\documentclass[useAMS,usenatbib,usegraphicx]{mn2e}
\bibpunct[,]{(}{)}{;}{a}{}{}

\usepackage{amssymb}

\newcommand{\hm}{h^{-1}}

\title[Galaxy population in Proto-Clusters]
{Semi-Analytic Model Predictions of the Galaxy Population in Proto-clusters}
\author[E.~Contini et al.]
        {E.~Contini,$^{1}$\thanks{Email: contini@pmo.ac.cn} 
         G.~De Lucia,$^{2}$
         N.~Hatch,$^{3}$
         S.~Borgani,$^{2,4,5}$
         X.~Kang,$^{1}$
         \\        
         $^1$ Purple Mountain Observatory, the Partner Group of MPI f\"{u}r Astronomie, 2 West Beijing Road, Nanjing 210008, China \\
         $^2$ INAF - Astronomical Observatory of Trieste, via G.B. Tiepolo 11, I-34143 Trieste, Italy \\
	 $^3$ School of Physics and Astronomy, University of Nottingham, University Park, Nottingham NG7 2RD \\
 	 $^4$ Dipartimento di Astronomia, Universit{\' a} di Trieste, via G.B. Tiepolo 11, I-34131 Trieste, Italy \\
 	 $^5$ INFN, Sezione di Trieste, Via Valerio 2, I-34127 Trieste, Italy \\}

\pagerange{\pageref{firstpage}--\pageref{lastpage}}
\pubyear{2015}

\begin{document}

\maketitle

\label{firstpage}

\begin{abstract} 
We investigate the galaxy population in simulated proto-cluster regions using a
semi-analytic model of galaxy formation, coupled to merger trees extracted from
N-body simulations. We select the most massive clusters at redshift $z=0$ from
our set of simulations, and follow their main progenitors back in time. The
analysis shows that proto-cluster regions are dominated by central galaxies and
their number decreases with time as many become satellites, clustering around
the central object.  In agreement with observations, we find an increasing
velocity dispersion with cosmic time, the increase being faster for
satellites. The analysis shows that proto-clusters are very extended regions,
$\gtrsim 20 \, Mpc$ at $z \gtrsim 1$. The fraction of galaxies in
  proto-cluster regions that are not progenitor of cluster galaxies varies with
  redshift, stellar mass and area considered. It is about 20-30 per cent for
  galaxies with stellar mass $\sim 10^9\,{\rm M}_{\sun}$, while negligible for
  the most massive galaxies considered. Nevertheless, these objects have
properties similar to those of progenitors.  We investigate the building-up of
the passive-sequence in clusters, and find that their progenitors are on
average always active at any redshift of interest of proto-clusters. The main
mechanism which quenches their star formation is the removal of the hot
gas reservoir at the time of accretion. The later galaxies are accreted (become
satellite), and the more the cold gas available, the longer the time spent as
active. Central galaxies are active over all redshift range considered,
although a non-negligible fraction of them become passive at redshift $z<1$,
due to strong feedback from Active Galactic Nuclei.
\end{abstract}


\section[]{Introduction} 
\label{sec:intro}

The term `proto-cluster' is used to refer to the over-density regions in the
early Universe that are believed to evolve into massive galaxy clusters today.
Putative proto-cluster regions were often localized around high-z radio
galaxies (HzRGs, e.g.,
\citealt{pentericci97,miley04,kurk04,venemans07,kuiper11}), that are among the
most massive galaxies at high redshift, and likely the progenitors of massive
elliptical galaxies residing at the centre of local massive clusters (e.g.,
\citealt{mclure99,zirm05,cooke08,miley08,hatch10}). Proto-cluster galaxies can
be efficiently identified using narrow-band imaging to detect emission line
(Ly$\alpha$ or H$\alpha$) objects at the redshift of the target
HzRG. Alternatively, broad-band imaging can be used, with colours chosen to
detect dropout objects at the target redshift. Follow-up spectroscopy is then
needed to confirm the redshift of the candidate proto-cluster galaxies. 
  Being proto-clusters regions of intense star formation activity, efficient
  searches can be conducted using the far-IR/submm bands. Recently,
  \cite{clements14} exploited the all sky coverage of the Planck satellite
  survey in combination with Herschel data, in order to detect candidate
  clusters undergoing dust-obscured violent star forming phase (see also
  \citealt{planck15}).

  While, for practical reasons, observational studies are often based on
  relatively small areas around putative sign-posts of proto-clusters
  \citep[e.g.][]{venemans07,hatch09}, it has been soon realized that these
  structures cover extended regions, up to $20 \, Mpc$
  \citep{kurk04,tanaka11,hatch10,toshikawa12}. Therefore, large areas are
  needed if the goal is to probe a large fraction of the proto-cluster galaxy
  population.

  From the theoretical viewpoint, a few recent studies have compared
  observational data with results from theoretical models of galaxy formation
  or used such models to interpret the observational
  results. \citet{Saro_etal_2009} compared results from hydrodynamical
  simulations of galaxy clusters to the observational properties of the
  Spiderweb galaxy system. Similar comparison work has been carried out more
  recently by \citet{granato15}, who pointed out that simulated cluster regions
  never reach the elevated star formation rates inferred from observational
  studies of proto-cluster regions. \citet{Overzier_etal_2009} used mock
  catalogues based on semi-analytic models of galaxy formation applied to the
  Millennium Simulation to study the relation between density enhancements
  around QSOs and proto-cluster regions at $z\sim 6$. More recently,
  \citet{chiang13}, \citet{orsi15}, and \citet{muldrew15} used similar
  techniques to study the relation between proto-clusters identified using
  common observational techniques and present day cluster descendants, and that
  between high redshift clusters and proto-clusters.

In this work, we adopt a similar approach and focus on the characterization of
the size and `contamination' from non cluster galaxies of proto-cluster
regions, star formation activity within the proto-clusters, and origin of the
passive sequence observed in local galaxy clusters. 

The paper is structured as follows: in Section \ref{sec:sim} we
present our set of simulations and the sample of proto-cluster
regions. Results from our case study are discussed in Section
\ref{sec:pcreg}, where we analyse the spatial distribution of central
and satellites galaxies in proto-clusters, their velocity dispersion
and total star formation rate. The latter will be discussed in detail
in Section \ref{sec:sfr} by using the full sample of proto-cluster
regions. In Section \ref{sec:stat_anal} we quantify the fraction of
progenitors in boxes centred around the central galaxy with different
sizes, to statistically characterize the typical size of proto-cluster
regions.  In Section \ref{sec:obs_pcr} we focus on the fraction of
outliers (defined as those objects that are not progenitors of
galaxies in cluster at $z=0$) in proto-clusters as a function of
galaxy stellar mass and redshift, while in Section \ref{sec:gal_popul} 
we follow the history of the proto-cluster galaxy population, focusing 
mainly on the progenitors of passive galaxies at $z=0$. Finally, we 
discuss our results and give our conclusions in Section \ref{sec:conclusions}.

\section[]{The simulated clusters}  
\label{sec:sim}

In this study we use N-body simulations of galaxy clusters,
generated using the `zoom' technique \citep*[][ see also
  \citealt{Katz_and_White_1993}]{Tormen_etal_1997}: a target cluster
is selected from a parent low-resolution simulation 
of a large cosmological volume and all its particles, as
well as those in its immediate surroundings, are traced back to their
Lagrangian region and replaced with a larger number of lower mass
particles. Outside this high-resolution region, particles of
increasing mass are displaced on a spherical grid. All particles are
then perturbed using the same fluctuation field used in the parent
cosmological simulations, but now extended to smaller scales. The
method allows the computational effort to be concentrated on the
cluster of interest, while maintaining a faithful representation of
the large scale density and velocity. All the cluster re-simulations used
  in this study are based on the same parent simulation. This followed $1024^3$
  dark matter particles within a box of $1 \hm$ Gpc comoving on a side.

Below, we use cosmological N-body simulations of 27 regions
surrounding as many massive clusters identified at z=0, and carried
out assuming the following cosmological parameters: $\Omega_m=0.24$
for the matter density parameter, $\Omega_{\rm bar}=0.04$ for the
contribution of baryons, $H_0=72\,{\rm km\,s^{-1}Mpc^{-1}}$ for the
present-day Hubble constant, $n_s=0.96$ for the primordial spectral
index, and $\sigma_8=0.8$ for the normalization of the power
spectrum. The latter is expressed as the r.m.s. fluctuation level at
$z=0$, within a top-hat sphere of $8\,\hm$Mpc radius. For all
simulations, the mass of each dark matter particle in the high
resolution region is $10^8\,\hm {\rm M}_{\odot}$, and the
Plummer-equivalent softening length is fixed to $\epsilon=2.3 \hm$~kpc
in physical units at $z<2$, and in comoving units at higher redshift.

For each simulation, outputs have been stored at 93 redshifts, between
$z=60$ and $z=0$. Dark matter haloes have been identified using a
standard friends-of-friends (FOF) algorithm, with a linking length of
0.16 in units of the mean inter-particle separation in the
high-resolution region. The algorithm {\small SUBFIND}
\citep{Springel_etal_2001} has then been used to decompose each FOF
group into a set of disjoint substructures, identified as locally
overdense regions in the density field of the background halo. Only
substructures that retain at least 20 bound particles after a
gravitational unbinding procedure are retained as genuine
substructures. Finally, merger histories have been constructed for all
self-bound structures in our simulations, using the same
post-processing algorithm that has been employed for the Millennium
Simulation \citep{springel05}. For more details on the simulations, as
well as on their post-processing, we refer the reader to
\citet{myself}. For our analysis, we use a sample of 27 haloes,
extracted from the high resolution regions of these simulations, and
with mass larger than $\sim 5 \cdot 10^{14} \hm M_{\odot}$. In
particular, 5 of our simulated haloes have $M_{200}\sim 10^{14}
M_{\odot}$, 9 $M_{200} \sim 7-8 \cdot 10^{14} M_{\odot}$ and the remaining
13 $M_{200} \sim 10^{15} M_{\odot}$.

We make use of the merger-trees extracted from our set of simulated clusters to
construct a mock catalogue of proto-cluster regions. The evolution of the
galaxy population is described by a semi-analytic model that is based on a
modified version of that presented in \citet{dlb}. In particular, we use the
updates presented in \cite{myself2}, that include an explicit modelling for the
formation of the intra-cluster light (ICL) via stripping processes and
mergers. We use the combination \emph{Model Tidal Radius+Mergers} presented in
that paper, but we stress that the results presented below are not affected by
the particular model used for the formation of the ICL. The updated model
  introduced in \cite{myself2} adopts the same parameter set used by the
  orginal model described in \cite{dlb}. The model was originally calibrated to
  match primarily the local K-band luminosity function. No high redshift data
  was used to tune the model parametes.

We run the semi-analytic model on the merger-trees extracted from the
simulations and generate galaxy-trees, i.e catalogues that contain,
for each model galaxy, information about all its progenitors and
descendants. We then select all galaxies within the virial radius
$R_{200}$ of each simulated cluster, and for each of them, we follow
its tree by considering all progenitors with stellar mass larger than
$2 \cdot 10^{8} M_{\odot}$ (this roughly corresponds to our resolution
limit). At each redshift, we consider the region occupied by all
progenitors of $z=0$ cluster galaxies as the proto-cluster region,
i.e. the boundaries of the proto-cluster regions are given by the
distribution of progenitors they contain. The center of each region is
defined by the position of the main progenitor of the central object 
found at redshift $z=0$. For the analysis that follows, unless otherwise 
stated, scale-lengths are given in comoving units.

In the following analysis, we have avoided including galaxies from the
low-resolution regions of each re-simulation. To this aim, when the
  analysis requires the selection of galaxies within some distance from the
  central galaxy, we have only considered galaxies in cubic regions of $\sim
7\, \hm Mpc$ on a side for the 5 clusters in our sample with mass $\sim
10^{14}\, \hm M_{\odot}$. These 5 clusters are not included in some analysis
considering larger regions of the proto-clusters (e.g. Figs.~\ref{fig:nobj_z}
and \ref{fig:nreal} below). For all other simulated clusters, the
high-resolution regions extends at least out to $\sim 15\, \hm Mpc$.

\section{The origin of a massive cluster}
\label{sec:pcreg}

\begin{center}
\begin{figure*}
\begin{tabular}{cc}
\includegraphics[scale=.45]{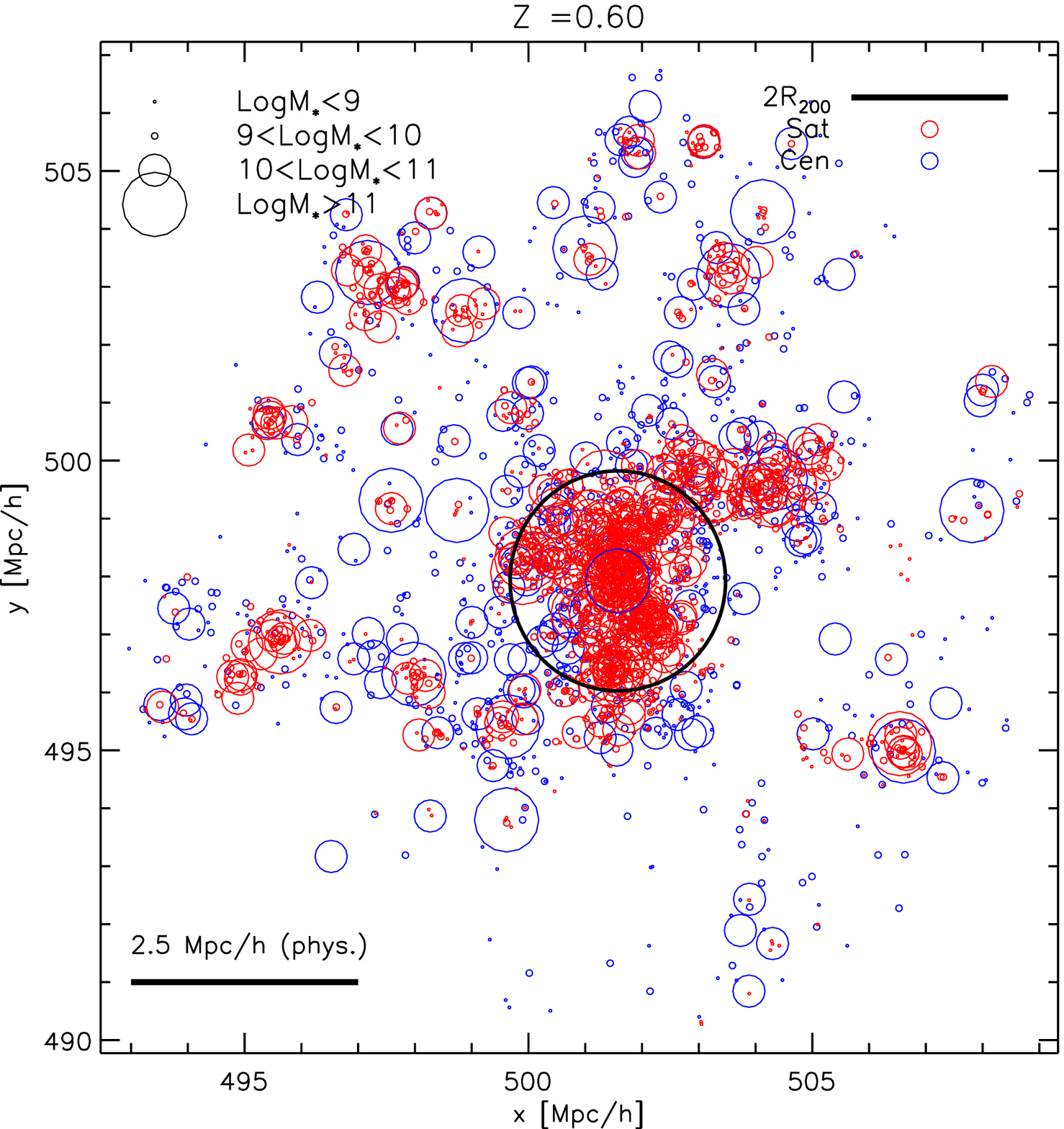} &
\includegraphics[scale=.45]{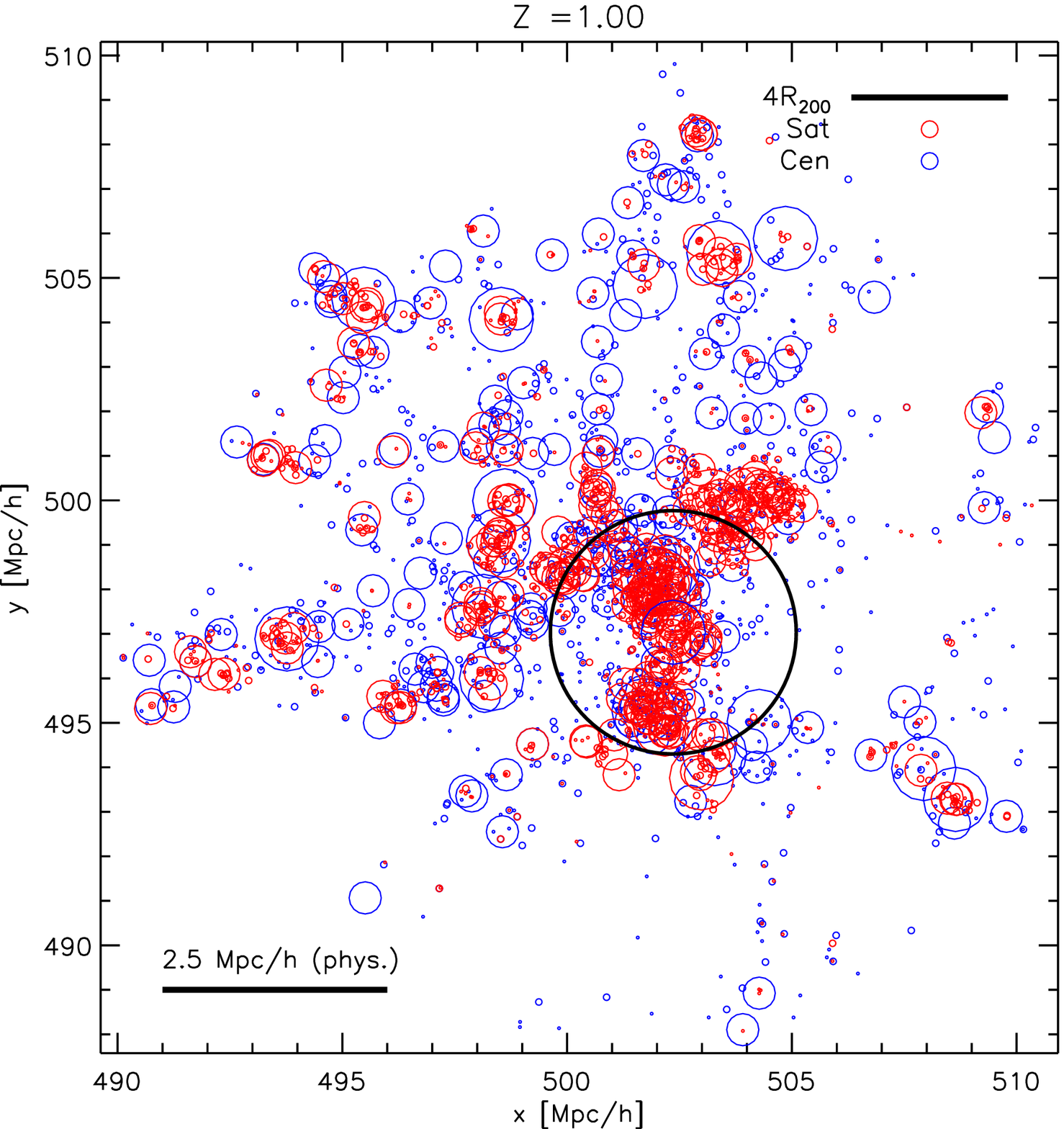} \\
\includegraphics[scale=.45]{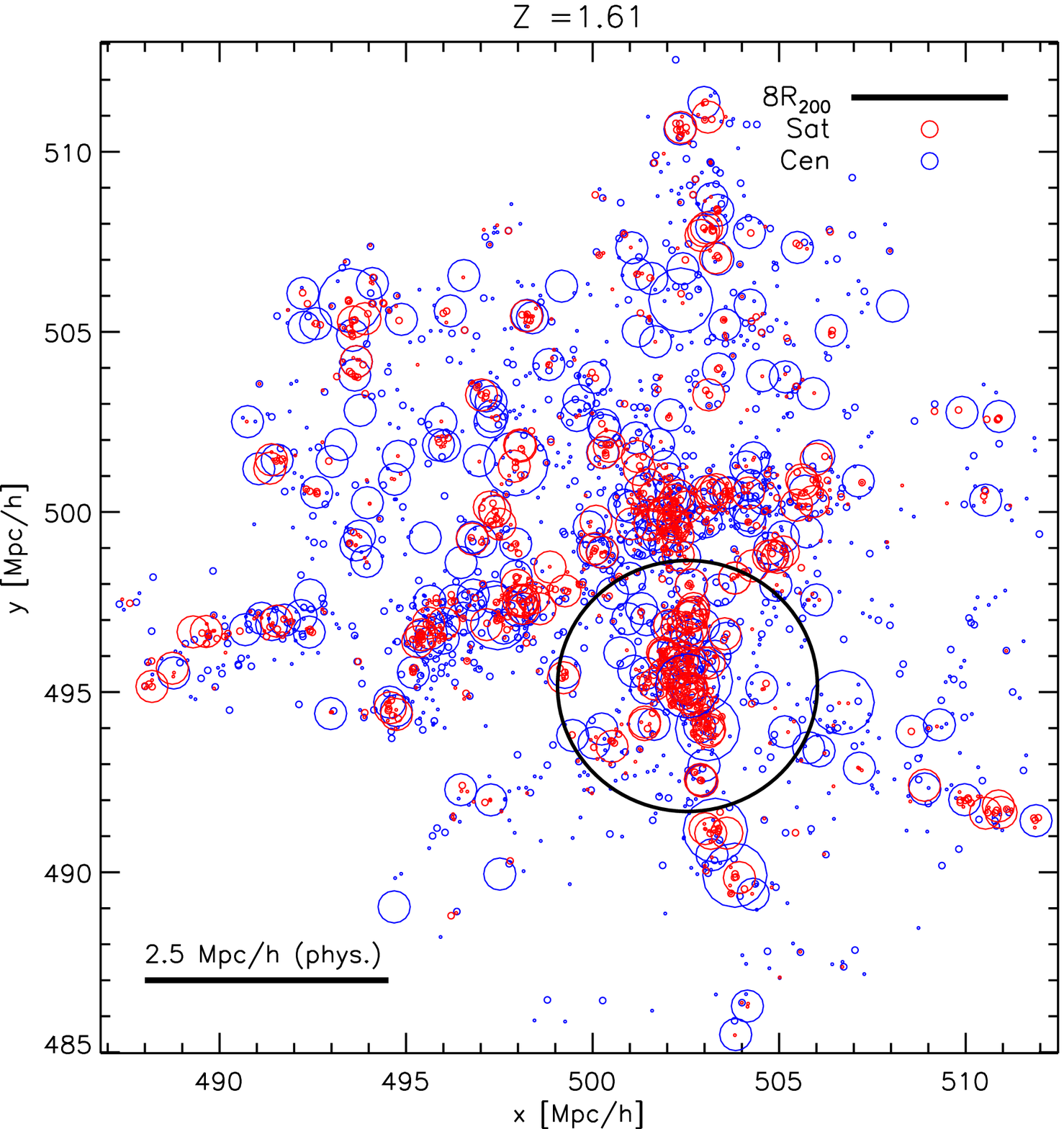} &
\includegraphics[scale=.45]{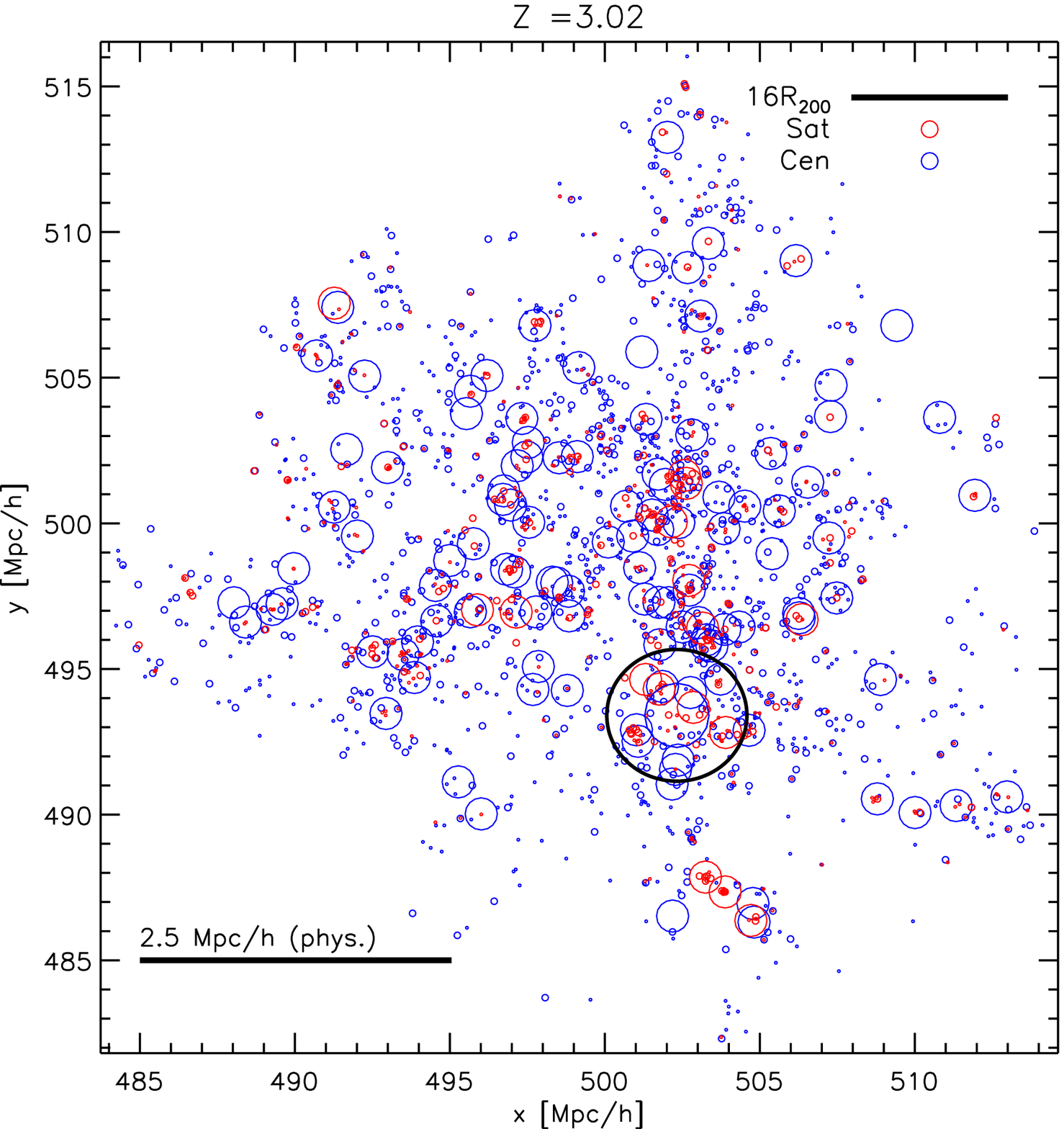} \\
\end{tabular}
\caption{X-Y positions of all progenitors of galaxies residing within
  $R_{200}$ of the most massive cluster in our sample at
  $z=0$. Progenitor distributions are plotted at four different
  redshifts, with red circles marking satellite galaxies and blue
  circles marking central galaxies. The dimension of each circle is
  proportional to the stellar mass of the galaxy. The black circle in
  each panel show a multiple of the virial radius of the halo
  containing the central galaxy at each z. The latter is identified as
  the galaxy sitting at the centre of the most massive cluster
  progenitor at each redshift.}
\label{fig:cl_pos}
\end{figure*}
\end{center}

\begin{center}
\begin{figure*}
\begin{tabular}{cc}
\includegraphics[scale=.40]{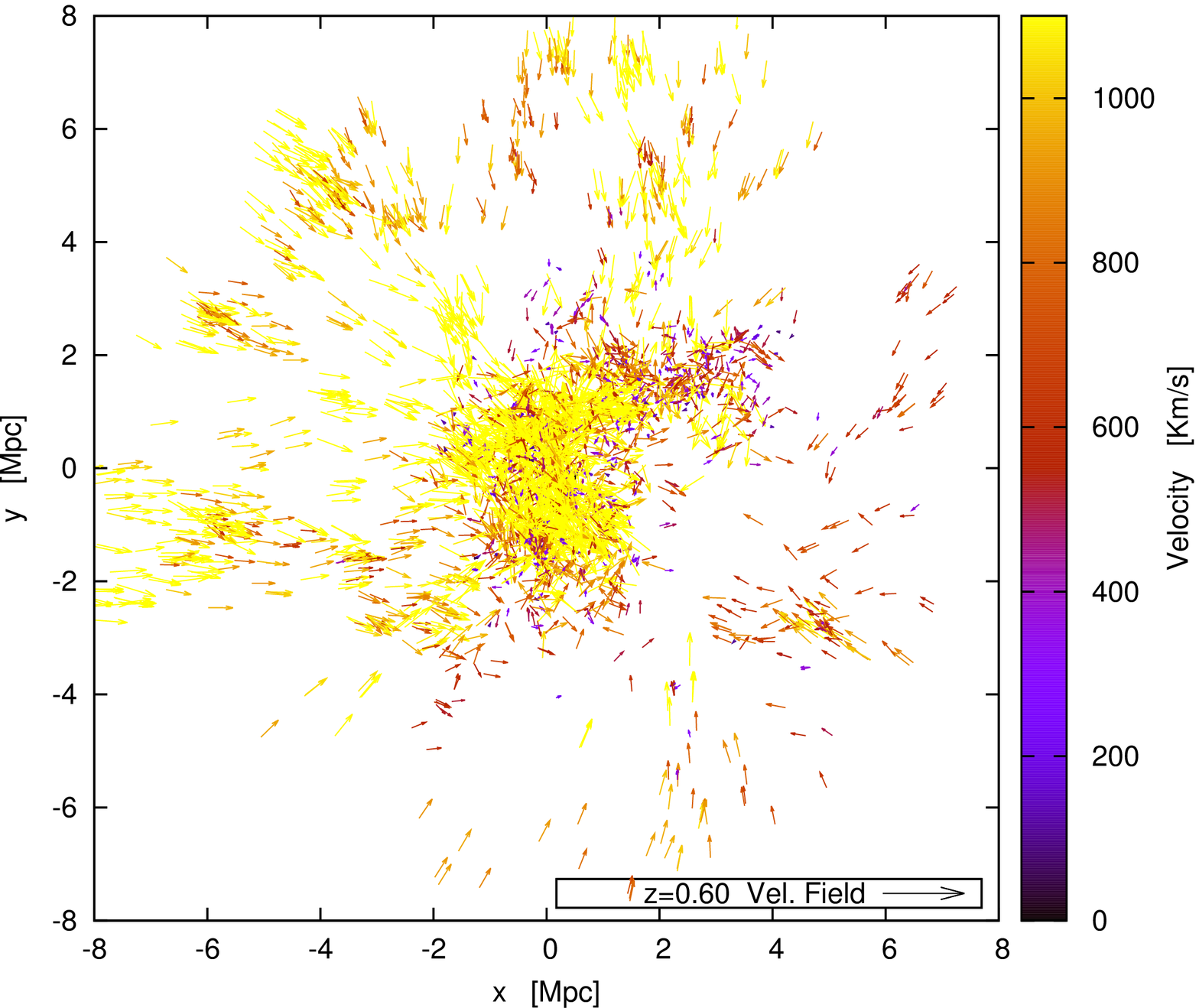} &
\includegraphics[scale=.40]{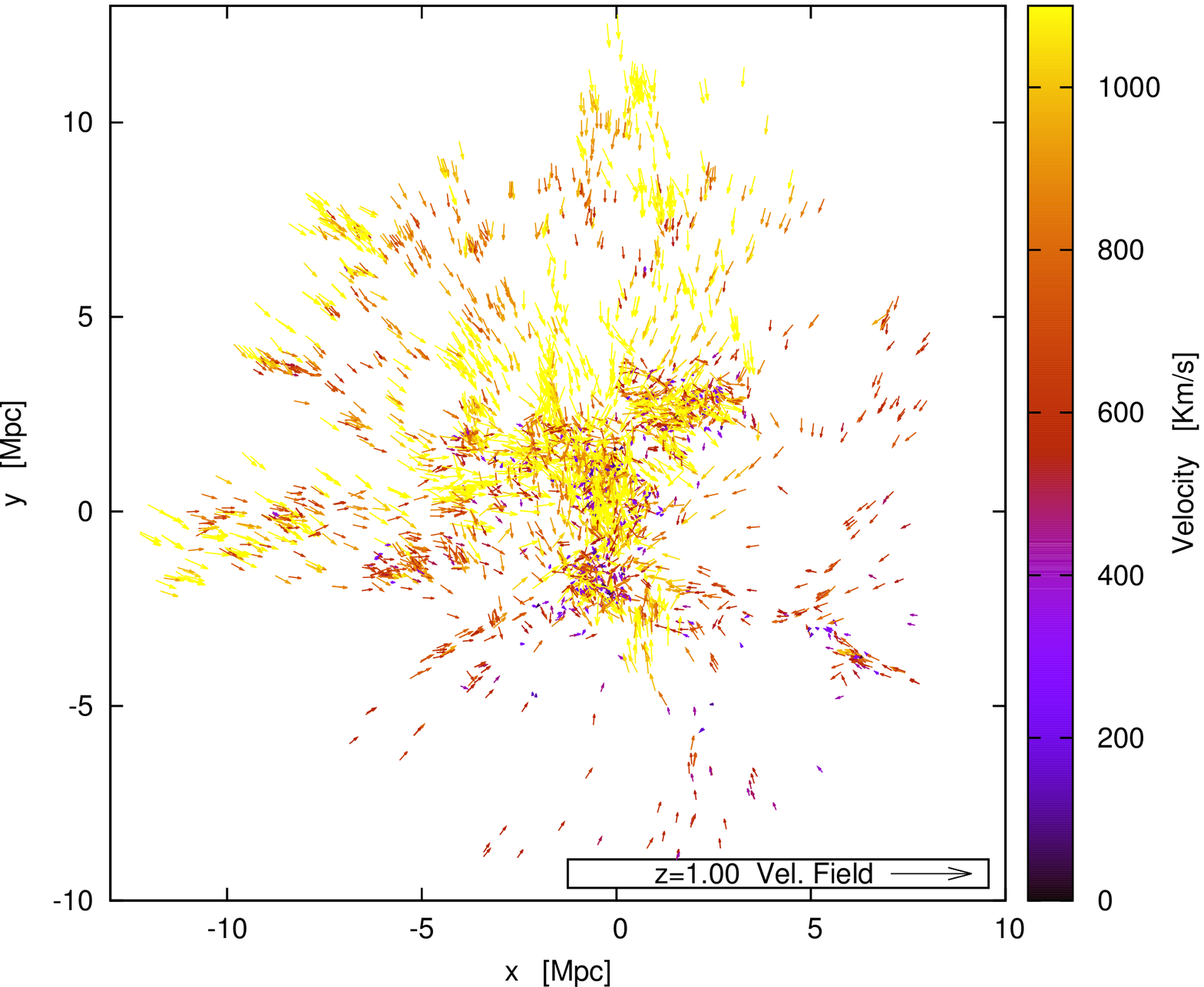} \\
\includegraphics[scale=.40]{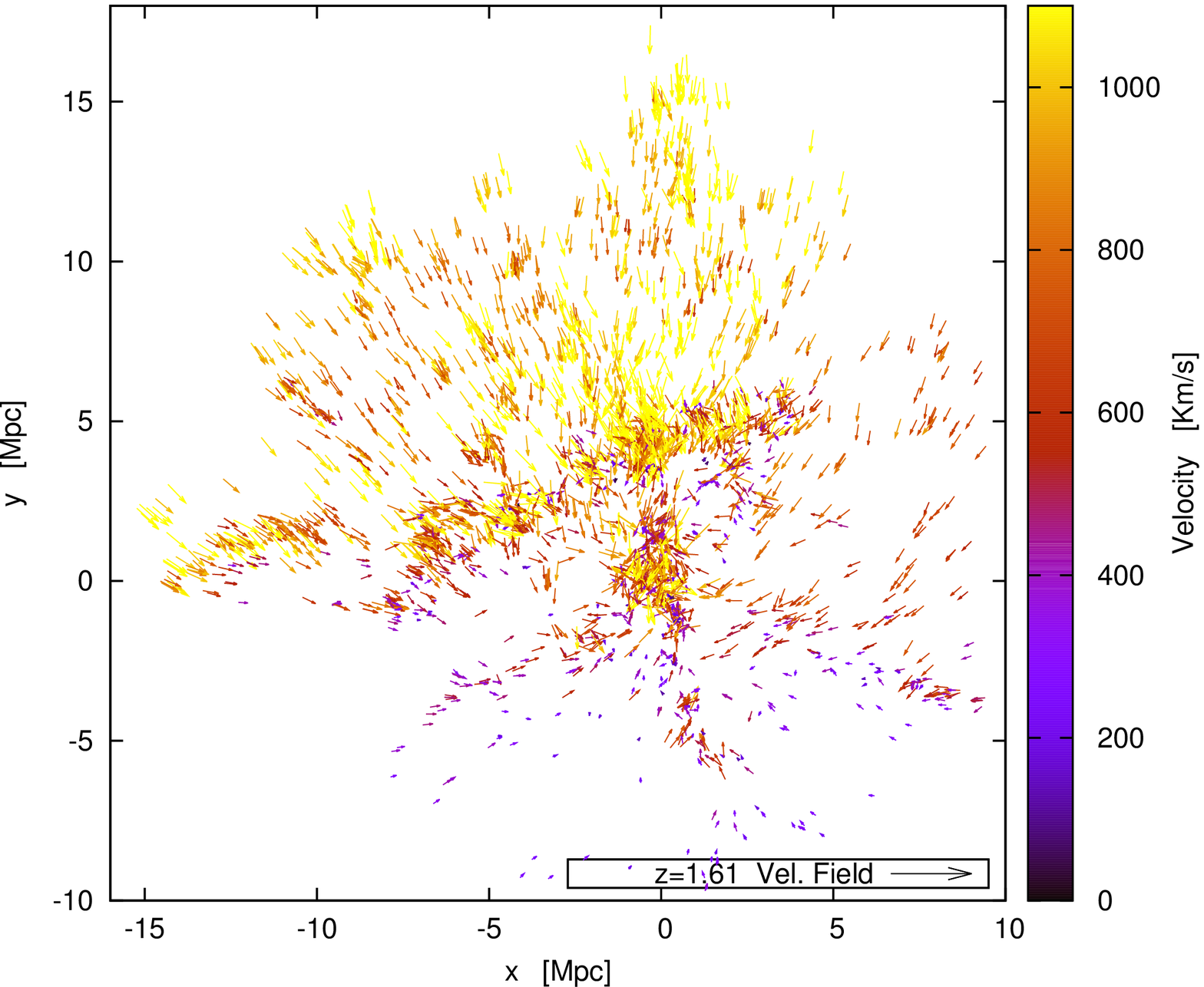} &
\includegraphics[scale=.40]{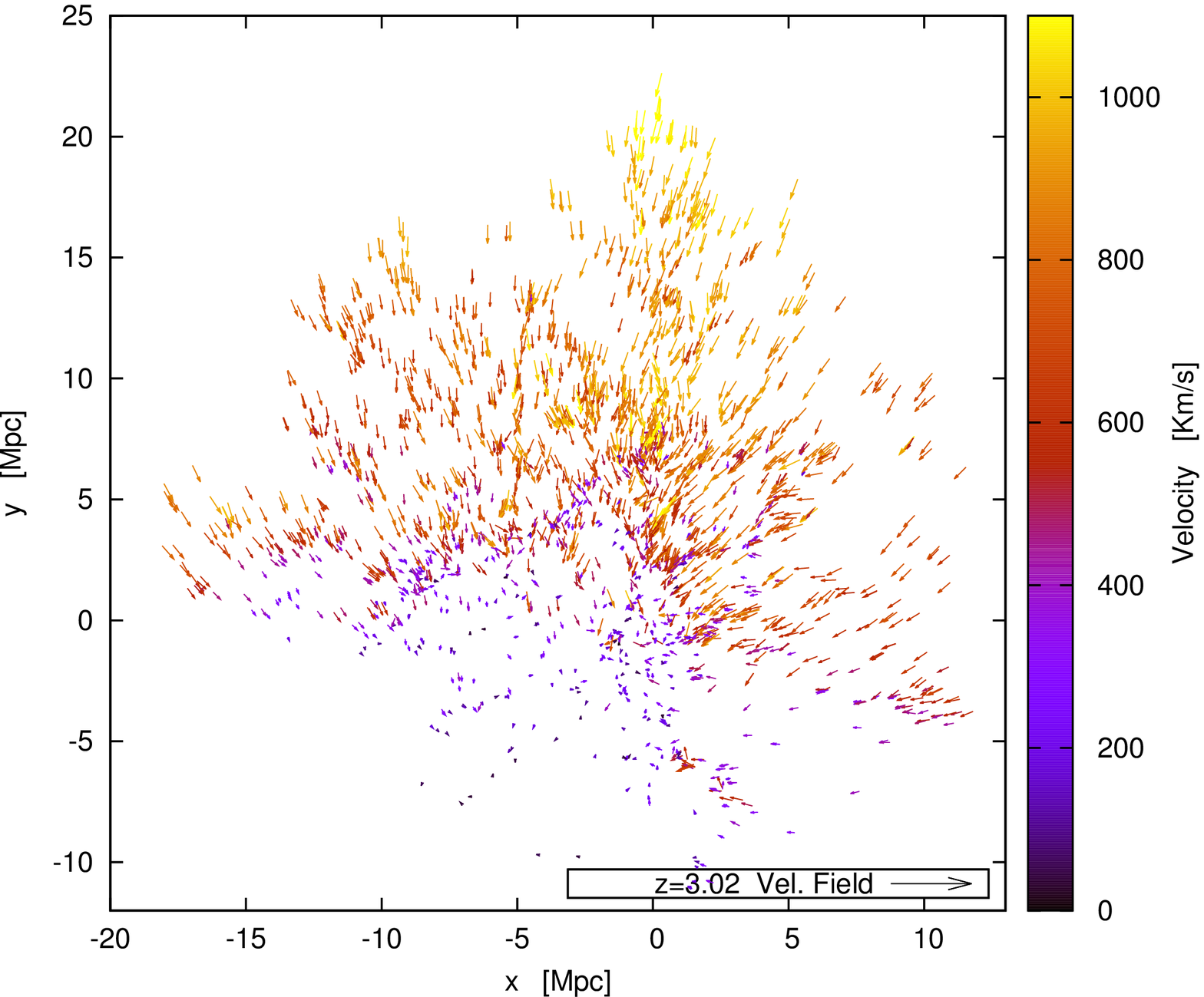}  \\
\end{tabular}
\caption{Velocity field of all progenitors of galaxies in the most
  massive cluster in our sample, at the same redshifts considered in
  Fig.~\ref{fig:cl_pos}. Positions and velocities have been computed 
  using all three components for each model galaxy, and shifting them with 
  respect to those of the CG, at each redshift.  
  Arrows indicate the direction of the motion while the modulus 
  is given by the colour coding.}
\label{fig:cl_vel}
\end{figure*}
\end{center}

\begin{figure*}
\begin{center}
\includegraphics[scale=.80]{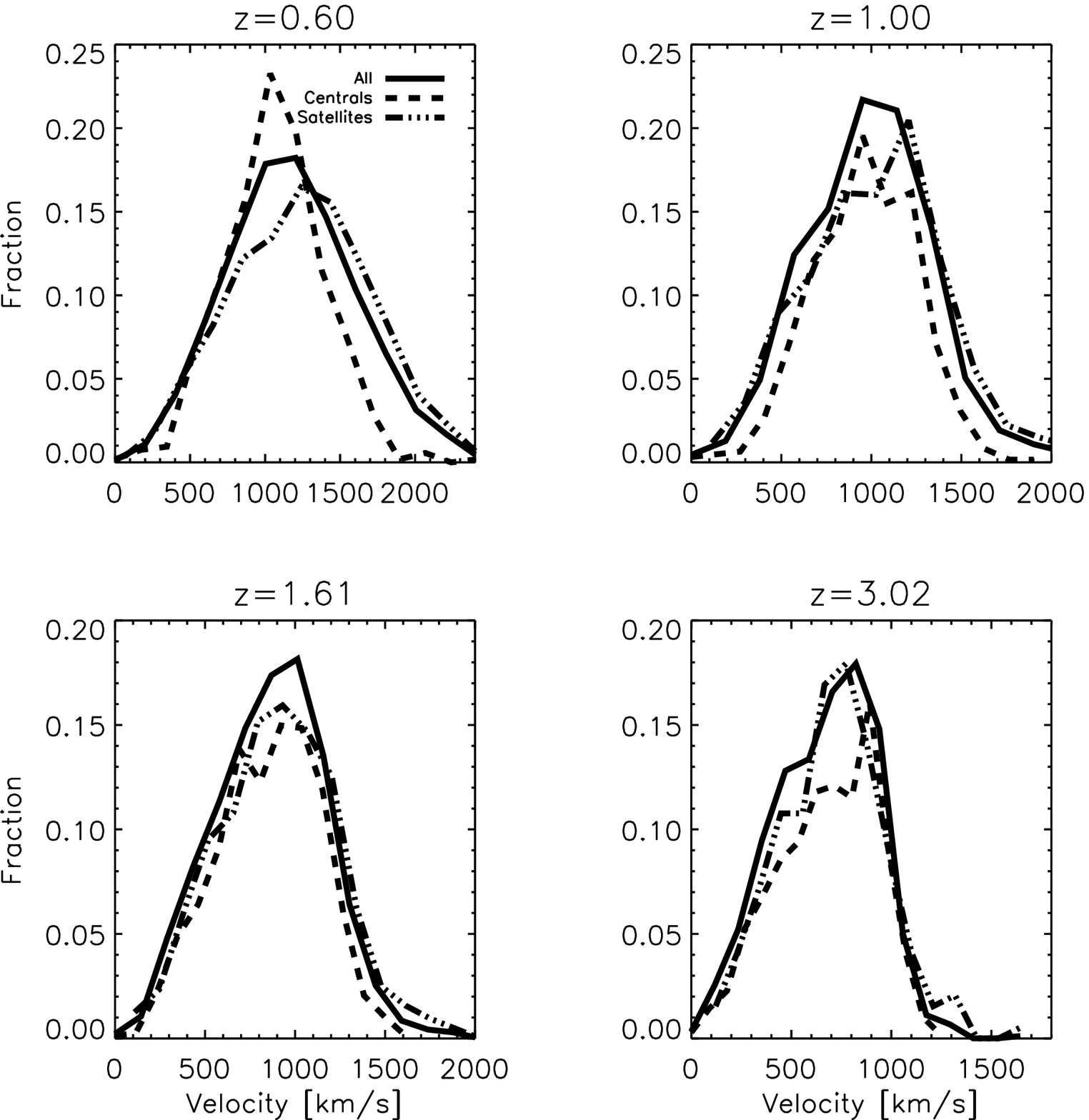} 
\caption{Velocity distribution for all (solid lines), central (dashed
  lines) and satellite (dash-dotted lines) progenitor galaxies, at the
  same redshifts considered in Figures \ref{fig:cl_pos} and
  \ref{fig:cl_vel}. Velocities have been computed using all three 
  velocity components for each model galaxy, and shifting them with respect 
  to those of the CG, at each redshift.}
\label{fig:veldistr}
\end{center}
\end{figure*}
In this section, we analyse in detail the proto-cluster region
corresponding to our most massive ($M_{200}\sim 10^{15} \hm
M_{\odot}$) galaxy cluster at $z=0$. In Figure \ref{fig:cl_pos}, we
show the x-y projections of progenitor positions at four different
redshifts: $z=0.60$, $1.00$, $1.61$, $3.02$. Satellite galaxies are
plotted in red, while centrals are in blue. The black circle in each
panel indicates a multiple of the virial radius (as indicated in the
legend) of the halo that contains the central galaxy (CG, hereafter). 
The latter is identified as the central galaxy in the main progenitor 
of the final cluster at each redshift.

Figure \ref{fig:cl_pos} shows that our most massive proto-cluster region is
dominated by central galaxies at high redshifts (see also \citealt{diener15}),
and that their number decreases with decreasing redshift. At redshift $z=0.60$,
most of the centrals have become satellites and entered the innermost regions
of the proto-cluster. As we will see below, central galaxies are typically
  star-forming systems at high-redshift. The fraction of central galaxies that
we find at high redshift ($\sim 0.7$) is consistent with recent results by
\citet{hatch10}, who find that $77 \pm 10$ per cent of galaxies in a sample of
proto-clusters at $2.2<z<2.6$ are blue. In our model, most of the satellites
are passive while centrals (excluded the most massive ones) are typically
active. At this redshift, we find that about 80 per cent of the galaxies in the
proto-cluster regions have specific star formation rate higher than $10^{-11}
\, yr^{-1}$. The comparison with results by \citet{hatch10} is just qualitative
at this stage. Their fraction is based on a sample of galaxies centred around 6
HzRGs and within a radius of 3.5 comoving Mpc, with redshifts between 2.28 and
2.55 from the collection of \cite{miley08}.  We will come back to this issue
below.

The CG grows by a factor four in stellar mass between $z \sim 3$ and $z \sim
1$, and then by a factor 2-3 between $z\sim 1$ and the present time
\citep{dlb,myself2}. The growth of the CG is mainly driven by accretion of
lower mass galaxies with the region around the CG not containing many galaxies
with comparable stellar mass over the redshift range considered. 
  \citet{hatch09} studied the stellar mass assembly of MRC 1138-262, also known
  as the Spiderweb Galaxy, a massive radio galaxy in a proto-cluster region at
  $z=2.2$. They identify the galaxies at the same redshift and within a
  projected distance of $150 \, kpc$ from the radio galaxy. Assuming that these
  satellites all lie on circular orbits around the radio galaxy, with radii
  given by their projected radii, they estimate their merging time-scale
  analytically (see their Eq. 3), and predict that most of them will merge with
  the central radio galaxy before $z=0$ increasing their mass by up to a factor
  2. In our test case, we find that 95 per cent of the satellites within
  $150\,kpc$ from the central galaxy will merge with it by $z=0$, in agreement
  with the calculation by Hatch et al.

In Figure \ref{fig:cl_vel}, we show maps of the velocity field at the
four redshifts considered for Fig.~\ref{fig:cl_pos}. Arrows indicate
the direction of the motion, and are colour coded accordingly to the
velocity modulus. Positions and velocities have been normalized to the
CG position and velocity at each redshift. At high redshifts,
progenitors have, on average, low velocities with respect to the
CG. This is because the halo potential is not very deep and the
progenitors are still relatively far away from it. The velocities tend
to increase towards $z=0$, as progenitors approach the central regions
of the cluster main progenitor. The halo potential also becomes deeper
with decreasing redshift as the cluster grows in mass.
 
In Figure \ref{fig:veldistr}, we plot the velocity distribution of
central (dashed lines), satellite (dash-dotted lines), and all
progenitor galaxies (solid lines) at the same redshifts of figures
\ref{fig:cl_pos} and \ref{fig:cl_vel}. The figure clearly shows that
the velocity distribution of centrals and satellites widen approaching
the present time, i.e. the velocity dispersion increases with cosmic
time, in good agreement with observations (e.g.,
\citealt{venemans07}) \footnote{We note that our velocity dispersions
  have been calculated using dynamical information provided by the
  simulation, while observationally they are determined using redshift
  information, and are typically based on a few proto-cluster members over
  relatively limited regions}. 

\begin{table}
\caption{Velocity dispersions for the three samples of galaxies used
  in Figure \ref{fig:veldistr} at the four different redshifts
  considered.  Units are $km/s$ and velocities have been normalized to
  the CG velocity.}
\begin{center}
\begin{tabular}{llllll}
\hline
Sample & z=0.60 & z=1.00 &  z=1.61 & z=3.02 \\
\hline
All        & 432 & 355 & 309 & 250 \\
Centrals   & 330 & 286 & 283 & 247 \\
Satellites & 460 & 391 & 335 & 258 \\
\hline
\end{tabular}
\end{center}
\label{tab:tab1}
\end{table}

In Table \ref{tab:tab1} we list the velocity dispersions of centrals,
satellites and all progenitor galaxies, at the four redshifts
considered. We see that the velocity dispersions of satellites and
centrals are similar at high-z, but that of satellites increases
faster, becoming larger than the velocity dispersion of centrals at
low redshift. This is because galaxies grow in mass as centrals and
then join the densest regions as satellites. As haloes grow in mass,
the velocity dispersion of satellites within them increases.

Results presented above show that the proto-cluster region of the most
massive cluster in our sample is very extended, $\sim 20 \, \hm Mpc$,
and hosts mainly a population of central galaxies at high redshift. At
this redshift, central galaxies are actively forming stars so that the
proto-cluster is a region of intense star formation. For the
particular example shown in this section, our model predicts a total
star formation rate of about $500 \, M_{\odot}/yr$ in the very inner
region and at $z \sim 2.6$.

\section{Fraction of Progenitors in Proto-Cluster Regions}
\label{sec:stat_anal}

\begin{figure*}
\begin{center}
\begin{tabular}{cc}
\includegraphics[scale=.60]{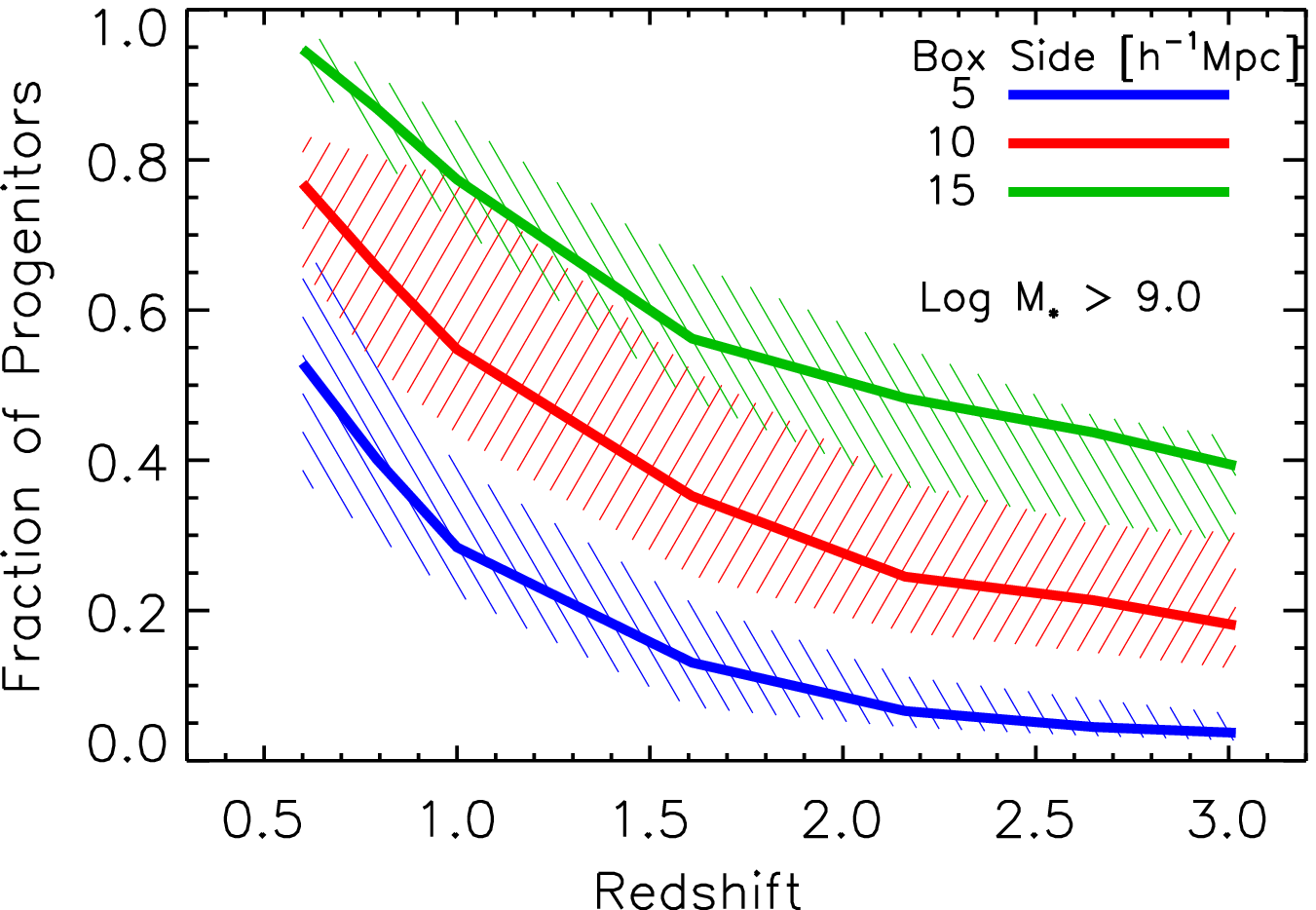} &
\includegraphics[scale=.61]{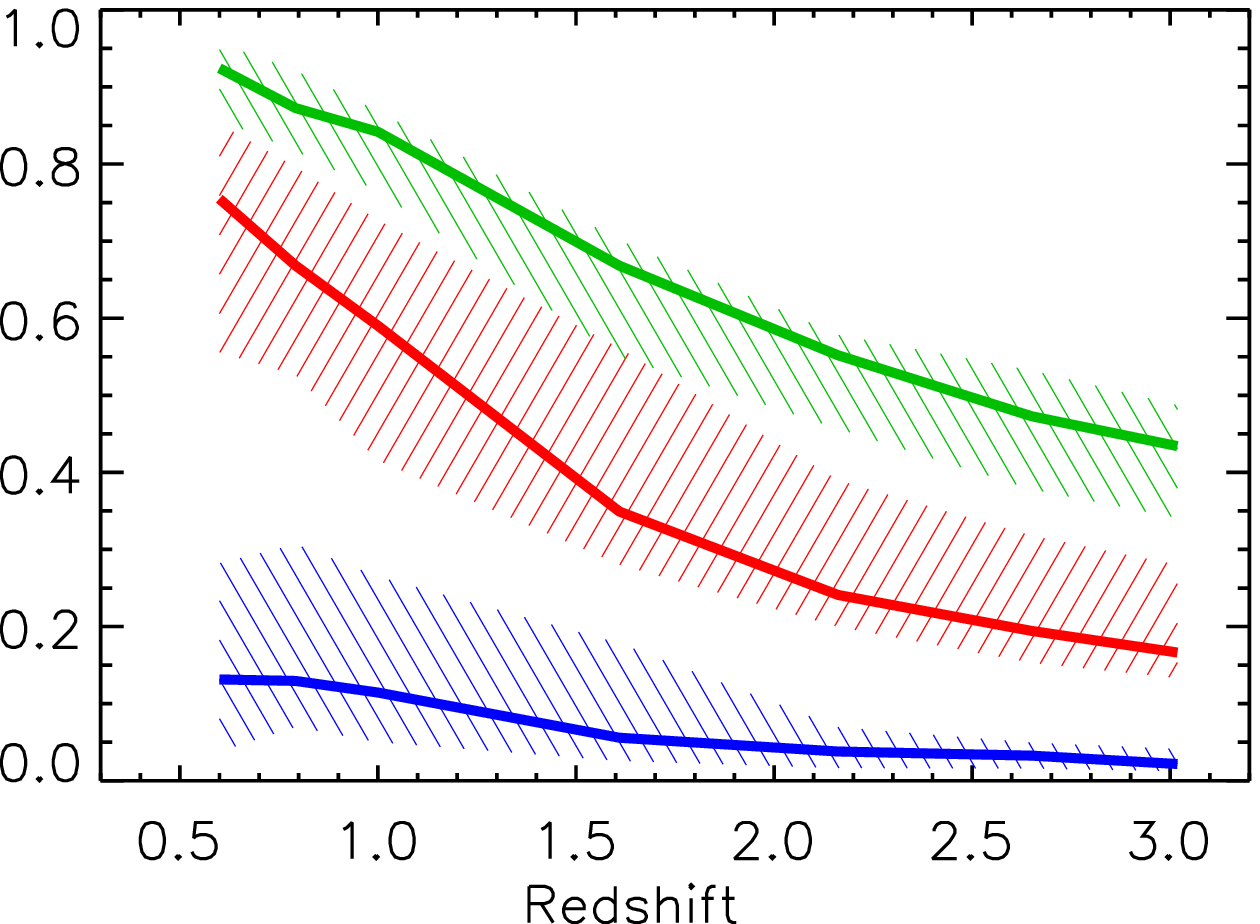} \\
\end{tabular}
\caption{Left panel: ratio between the number of progenitors contained in cubic
  boxes of different sizes ($5, 10,15 \, \hm Mpc$) centred around the CG at
  each redshift, and the total number of progenitors in the proto-cluster
  region (including those outside the box) with stellar mass above $10^9
  M_{\odot}$, as a function of redshift. Solid lines represent the median
  relations, while shaded regions mark the 20th-80th percentiles. Fractions
  have been computed by considering the CG position as the centre at each
  redshift. Right panel: same as left panel, but computing fractions by
  considering the geometrical centre (median x, y, z of all proto-cluster
  galaxies at each redshift).}
\label{fig:nobj_z}
\end{center}
\end{figure*}

\begin{figure}
\begin{center}
\includegraphics[scale=.5]{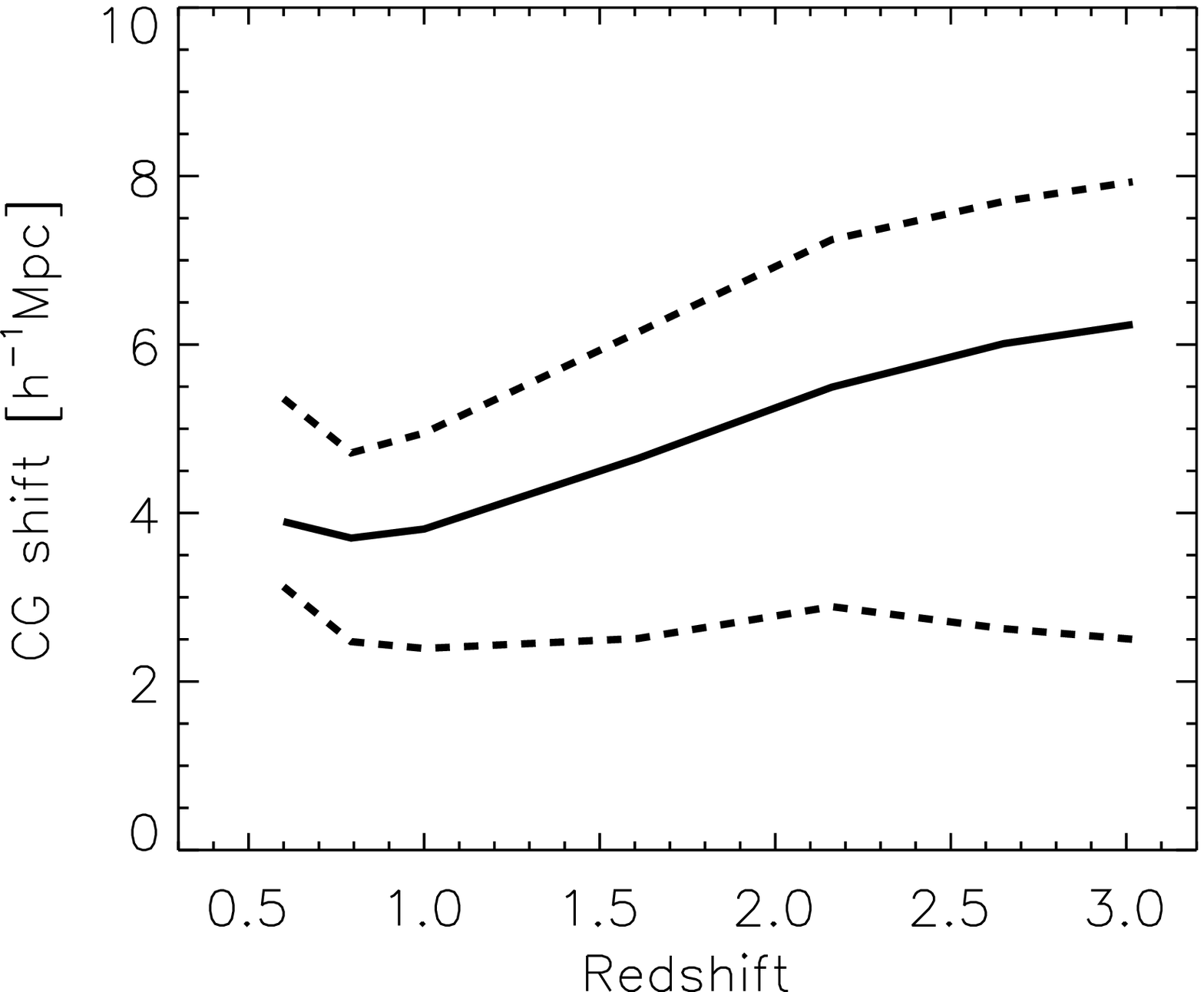} 
\caption{Shift of the CG with respect to the geometrical centre of the
  proto-cluster region, as defined in Figure \ref{fig:nobj_z}. Solid
  lines represent the median, while dashed lines represent the 10th
  and 90th percentiles of the distribution.}
\label{fig:cgshift}
\end{center}
\end{figure}

The results discussed in the previous section confirm that proto-clusters are
extended objects, and that their size clearly depends on the particular
redshift at which they are observed. In this section, we focus on the
characterization of the typical size of proto-cluster regions by quantifying
the fraction of progenitors in regions of different comoving size. The number
of systems in our sample (27) allows us to provide also an estimate of the
halo-to-halo scatter. Our results do not depend significantly on the mass of
the final cluster and, therefore, we discuss the average trend of our complete
sample. In a recent study, \citet{chiang13} find a tight correlation between
the size of the proto-cluster and the cluster final mass, which appears in
contradiction with our previous statement. We note, however, that
\citet{chiang13} identify proto-clusters using a different method and consider
a larger dynamical range in the mass of the final cluster. In line with
  our results are those by \cite{orsi15}, who provide a prediction for
  the evolution of a typical size of proto-clusters with redshift that does not
  depend significantly on final cluster mass.

In the left panel of Figure \ref{fig:nobj_z} we plot the ratio between
the number of progenitors contained in cubic boxes of different sizes
($5, 10,15 \, \hm Mpc$) centred around the CG at each redshift, and
the total number of progenitors with stellar mass above $10^9
M_{\odot}$ (we do not find significant differences using higher cuts in
stellar mass) as a function of the proto-cluster redshift. Solid lines
show the median fractions, while shaded regions mark the 20th-80th
percentiles area. Proto-clusters are often identified around high-z
($z \gtrsim 1.5$) radio galaxies, considering areas of few $Mpc^2$
(typically less than $2 \times 2 \, Mpc$ physical) around the radio
galaxy. Over the same redshift range, for the proto-clusters
considered in our study, the fraction of progenitors varies between
0.2 to almost zero if computed in a box of $5 \, \hm Mpc$ (blue solid
line and shaded area), and between 0.6 to 0.4 within a box of $15 \,
\hm Mpc$ (green solid line and shaded area). Therefore, as stressed
above, very large regions are needed in order to include the bulk of
the galaxy population in proto-clusters at high redshift.

The right panel of Figure \ref{fig:nobj_z} shows the same quantities
given in the left panel, but now considering the geometrical centre of
the proto-cluster region, defined by the median x, y, and z of all
proto-cluster galaxies. The fractions corresponding to the $10$ and
$15 \, \hm Mpc$ boxes are similar to those plotted in the left
panel. For the smallest box considered, the fractions computed around
the CG are about twice those computed using the geometrical centre for
$z > 2$. Therefore, the distribution of progenitors in the sky is not 
typically symmetric around the CG.

In Figure \ref{fig:cgshift}, we plot the median CG distance from the
geometrical centre (solid line) of the proto-cluster region as a function of
redshift. Dashed lines indicate the 10th and 90th percentiles of the
distributions. The shift between the CG position and the geometrical centre is
always larger than $\sim 2 \, \hm Mpc$. It becomes $\sim 5 \, \hm Mpc $ at $z
\sim 2$, and $\sim 6 \, \hm Mpc$ at $z \sim 3$. This is due in part to the
  high number of progenitors residing in haloes that have not yet merged with
  the main progenitor of the final cluster. In addition, the bulk motion of
  progenitors, as seen in Figure \ref{fig:veldistr}, causes a decrease with
  decreasing redshift of the shift.


\section{Outliers in Proto-Cluster Regions}
\label{sec:obs_pcr}

\begin{figure*}
\begin{center}
\includegraphics[scale=.85]{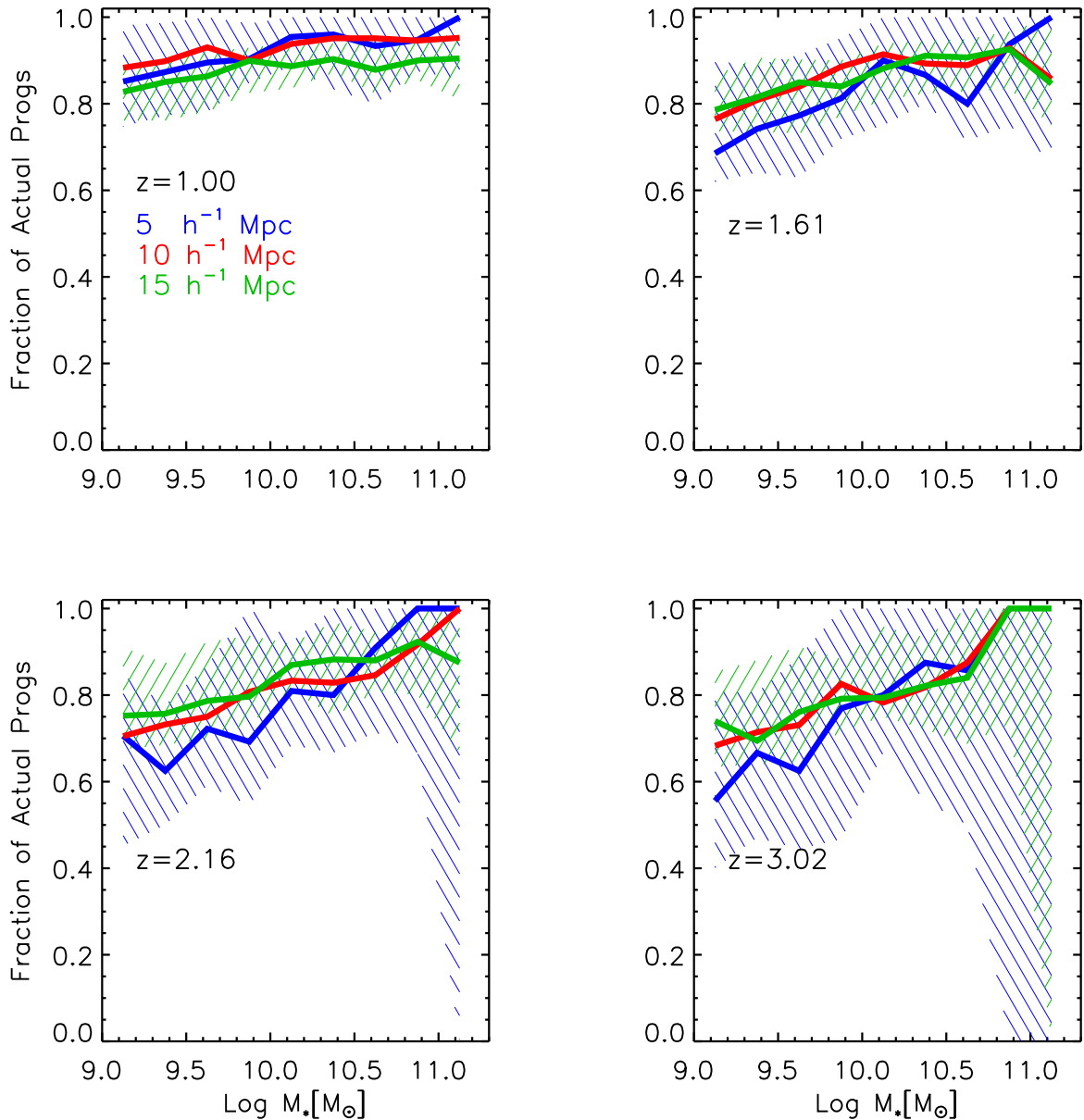} 
\caption{Fraction of actual progenitors in our simulated proto-cluster regions
  within cubic boxes of different sizes as indicated in the legend, as a
  function of stellar mass and at different redshifts. Solid lines represent
  the median, while shaded areas show the distribution between the 20th and
  80th percentiles for the smallest and largest boxes. The scatter around
    the red solid lines is comparable to that corresponding to the largest box
    size.}
\label{fig:nreal}
\end{center}
\end{figure*}

As mentioned above, proto-cluster regions are often identified around
luminous galaxies at high redshifts. Our simulated catalogues allow us
to estimate what is the typical fraction of `outliers', i.e. of
galaxies in the region that are not actual progenitors of a galaxy
residing in the descendant cluster at $z=0$. In particular, we will
consider `outliers' all galaxies that are not progenitors of cluster
galaxies located within $2 \cdot R_{200}$ of the cluster at $z=0$.

In Figure \ref{fig:nreal} we plot the fraction of actual progenitors
in our proto-cluster regions within cubic boxes of different sizes
($5$ for all 27 regions considered in 
this study, and $10, 15 \, \hm Mpc$ for the 22
regions corresponding to the 22 most massive clusters at $z=0$)
centred around the CG, as a function of their stellar mass, and at
four different redshifts. Solid lines represent the median fractions,
while shaded areas represent the distribution between the 20th and
80th percentiles. This scatter is larger at increasing redshift,
irrespective of the size of the box. The plot shows that the fraction
of actual progenitors weakly increases with progenitor stellar mass,
between 0.6 for progenitors with stellar mass around $M_* \sim 10^{9}
M_{\odot}$, and 1 for stellar masses of about $10^{11} M_{\odot}$ at
$z \gtrsim 2$. The trend is analogous at lower redshift, but the
fraction is higher at low stellar mass. The fraction at high stellar
mass and at high redshift is affected by low-number statistics.

Our analysis show that, in our simulated proto-cluster regions, the
fraction of outliers depends on the galaxy stellar mass and (weakly) on
the redshift at which proto-cluster regions are located, and does not
strongly depend on the size of the box up to 15 comoving Mpc. 
We find that about 30 per cent
of galaxies with stellar mass smaller than $\sim 10^{10} M_{\odot}$
are outliers at $z \sim 3$, and that this fraction rapidly decreases
towards larger stellar mass. Slightly smaller fractions are found at
lower redshifts, and when considering progenitors of galaxies within
$R_{200}$ (instead of $2 \cdot R_{200}$) at redshift zero.

Our analysis also shows that it is virtually impossible to distinguish between
outliers and actual progenitors by looking at their physical properties: we
have verified that the distribution of colours, star formation rates, cold, hot
and stellar masses or location in the proto-cluster regions for outliers do not
differ significantly from those of progenitors. We have verified that 
  the proto-cluster members and outliers also share similar line-of-sight
  velocity distributions. However, it is worth noting that this set of
simulations is not ideal for addressing this issue. In fact, the
high-resolution region of our cluster re-simulations only allow us to consider
regions of the Universe corresponding to the surroundings of massive
clusters. Therefore, these outliers may not be representative of the average
`field' galaxy.

Finally, we stress that we do not mimic exactly the typical observational
procedure that is adopted to select proto-cluster galaxy candidates. This would
require the construction of light-cones and the application of similar
selection procedures as done by other authors (e.g. \citealt{overzier08}.)
However, our results indicate that even with very high spectral resolution 
observations, we will be unable to distinguish outliers from true proto-cluster 
galaxies as they occupy the same volume.

\section{Star Formation Rate in Proto-Clusters}
\label{sec:sfr}

Proto-clusters are regions of strong star formation activity
(\citealt{pentericci01,miley06,overzier08,hatch08,hatch11,tanaka11,hayashi12}).
\citet{tanaka11} report the discovery of a significant excess, about a factor 5
with respect to the field, of candidate H$\alpha$ emitters in the proto-cluster
associated with the radio galaxy 4C 23.56 at $z=2.48$. Combined with
mid-infrared photometric data, they conclude that active star formation must be
occurring in the proto-cluster region around the radio galaxy, and that its
rate should be comparable with that of the average field at the same
redshift. \citet{hatch11}, who investigate the proto-clusters surrounding MRC
1138-262, at $z \sim 2.2$, and 4C + 10.48, at $z \sim 2.35$, find a total star
formation rate within the central $1.5 \, Mpc$ of about $ 5000 \, M_{\odot}/yr$
for the former, and about $3000 \, M_{\odot}/yr$ for the latter, much higher
than the typical star formation rate in local galaxy clusters
(\citealt{koyama10,wetzel12}, and references therein). \citet{hayashi12},
however, find smaller total star formation rates in three clumps (two of which
corresponding to areas larger than $4.5 \, Mpc^2$) around the radio galaxy USS
1558-003 at redshift $z=2.53$.  In the central region $50 \times 40 \, kpc$ of
MRC 1138-262, \citet{miley06} find a total star formation rate $\gtrsim 100 \,
M_{\odot}/yr$, in good agreement with the total star formation rate, $130 \pm
13 \, M_{\odot}/yr$, that \citet{hatch08} obtain in the central region $65
\times 65 \, kpc$ of the same object. A similar amount of star formation ($302
\, M_{\odot}/yr$) is found by \citet{hatch11} in the central region, $100
\times 100 \, kpc$, of $4C+10.48$. Very high `total' star formation rates
  ($\sim 10^4 {\rm M}_{\sun}/{\rm yr}$) have been measured recently by
  \citet{clements14} for overdensities of Herschel sources.

\begin{figure}
\begin{center}
\begin{tabular}{cc}
\includegraphics[scale=.42]{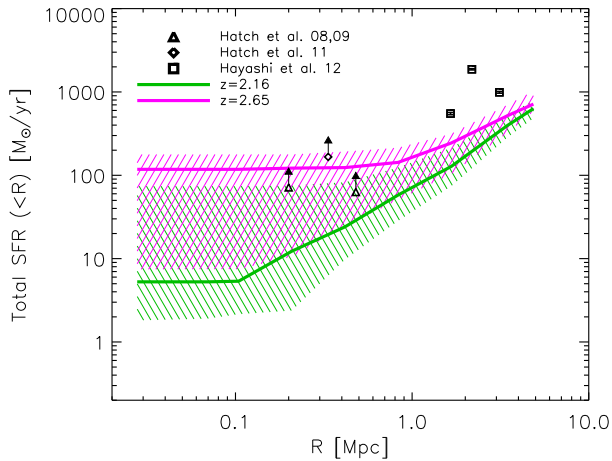}  
\end{tabular}
\caption{Total star formation rate (cumulative) of galaxies
  in the proto-cluster regions as a function of the projected distance
  from the centre, at different redshifts (lines of different
  colours). Solid lines represent the median value calculated on the
  27 proto-cluster regions, and shaded areas represent the
  distribution between the 20th and 80th percentiles. Black triangles
  and diamond represent the observational data by
  \citet{hatch08,hatch09}, and \citet{hatch11}, respectively, and
  black squares represent observational measurements by
  \citet{hayashi12}.}
\label{fig:totsfr}
\end{center}
\end{figure}

In Figure \ref{fig:totsfr} we show the total star
formation rate of galaxies in our proto-cluster regions within an area
of size plotted on the x-axis, at different redshifts (lines of
different colours). Solid lines represent the median value, while
shaded areas represent the distribution between the 20th and 80th
percentiles. The total star formation rate of galaxies within a given
area of our proto-cluster regions is a decreasing function of
redshift, in qualitative agreement with observations. Quantitatively,
however, our model predictions are offset with respect to the
observational data. In particular, the observational data by
\citet{hatch08} and \citet{hatch09} (black triangles) should be
compared with the green line, while \citet{hatch11} (black diamond),
and \citealt{hayashi12} (squares) should lie between the green and
magenta lines. 

All these studies (but \citealt{hatch09}) use a Salpeter IMF
(\citealt{salpeter55}), while our model (as \citealt{hatch09}) adopts a
Chabrier (\citealt{chabrier03}) initial mass function (IMF). Following
\citet{longhetti09}, we have corrected star formation rate estimates based
  on the assumption of a Salpeter IMF, using the following
  conversion 
\begin{displaymath}
SFR_{Cha}(z)=0.55 \cdot SFR_{Sal} (z).
\end{displaymath}
We also note that
the observational estimates are dust-uncorrected.  Assuming a minimum of dust
extinction, especially in the inner regions, they move up to the upper limits
of our predictions, even beyond $100 \, M_{\odot}/yr$.  Hence, despite the
large object-to-object scatter especially at high redshift and in the inner
regions, our model predicts star formation rates that are lower than
observational estimates by a factor of 2 in the innermost regions and up to a
factor of 5 or so at larger radii.  Interestingly, a recent work by
\citet{granato15} also finds too low SFRs in hydrodynamical simulations of
proto-clusters, when comparing theoretical results with observational estimates
by \citet{clements14}. It should be noted that most if not all proto-clusters
known at $z>2$ are selected as overdense regions around radio galaxies. The
elevated star formation rates measured for distant radio galaxies can only be
sustained for a short period of time. Therefore, radio galaxy selected
proto-clusters might be a special subset of proto-clusters with very high star
formation rates, which could at least in part explain the mismatch between data
and model predictions.

\section{Passive-Galaxy Sequence}
\label{sec:gal_popul}

Passive elliptical-like galaxies represent a significant fraction of
the local cluster galaxy population, particularly in the
high-mass end. The spectroscopy of these galaxies becomes difficult at
$z>1$ (\citealt{cimatti02,daddi04,mvw})
not only because of the scarcity of such objects, but also because of the 
intrinsic difficulty in detecting them (usually they require a good 
continuum signal-to-noise ratio) and measuring their redshifts (see e.g.
\citealt{gobat11}). However, recent observations demonstrate that the densest 
cores of most evolved cluster progenitors already host a (small) population 
of massive quiescent galaxies (see, e.g. \citealt{fassbender14,strazzullo15}, 
and references therein).

\begin{center}
\begin{figure*}
\begin{tabular}{cc}
\hspace{0.3cm} \includegraphics[scale=.48]{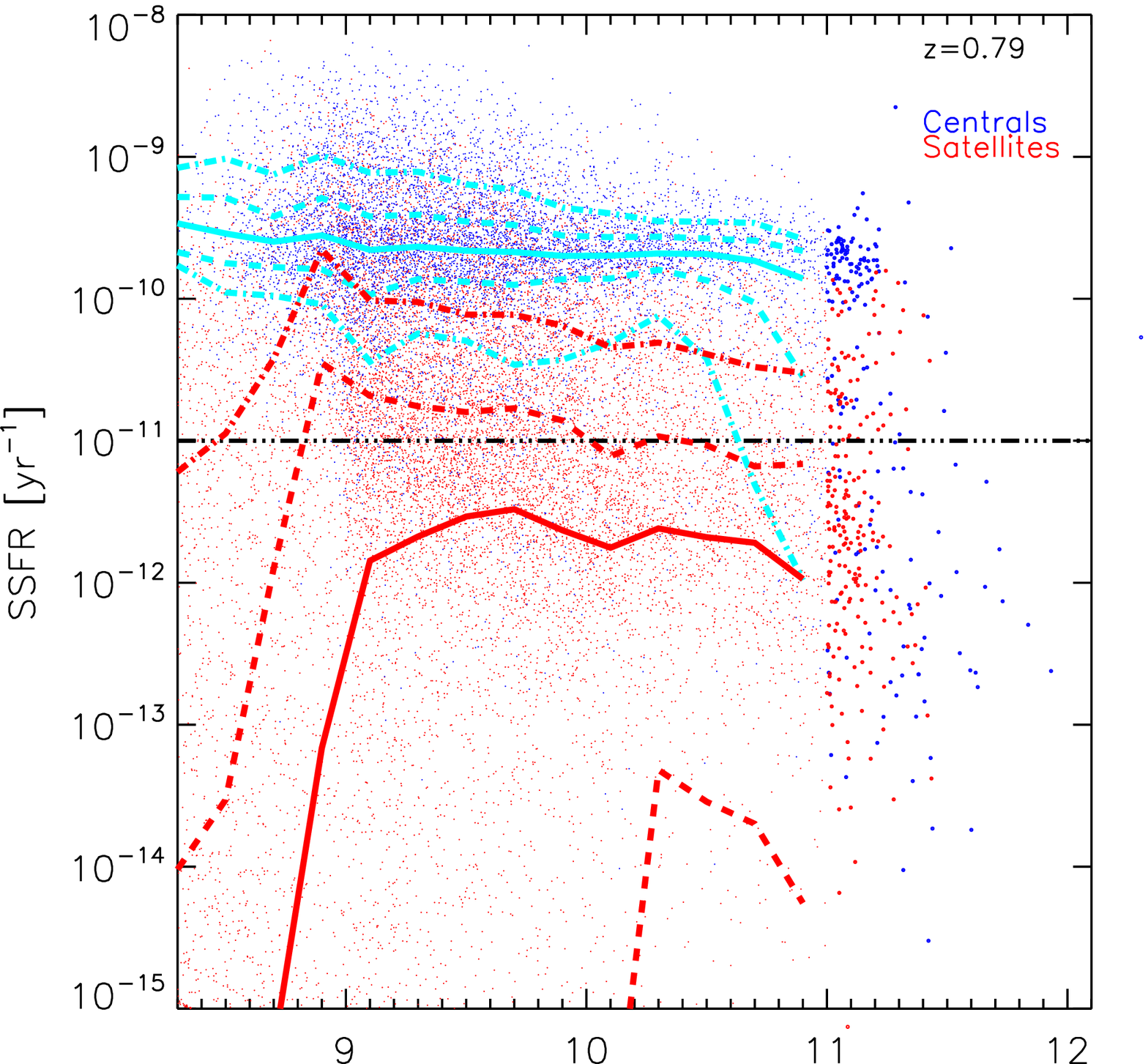} &  
\includegraphics[scale=.48]{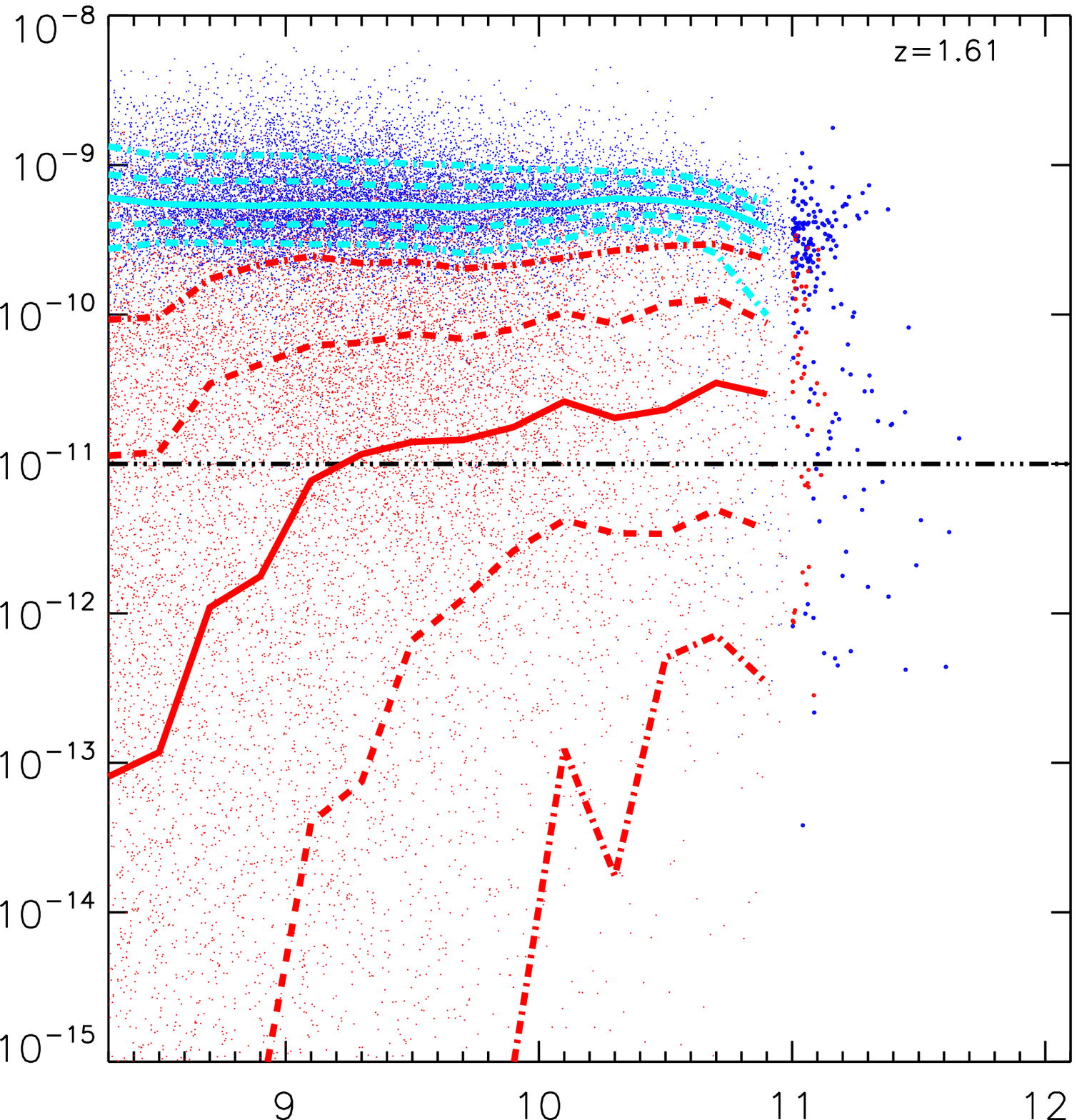} \\
\includegraphics[scale=.48]{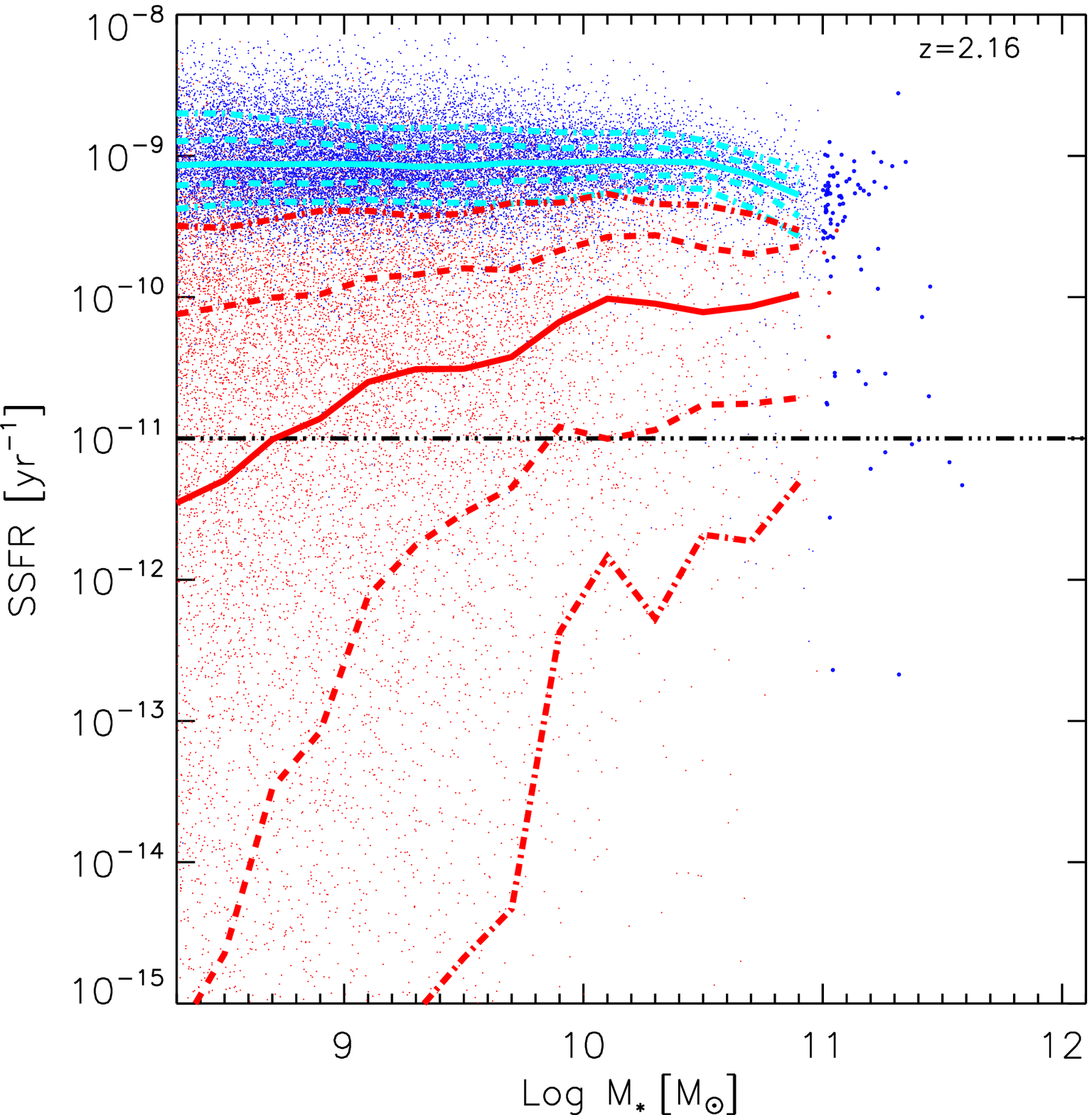} & 
\includegraphics[scale=.48]{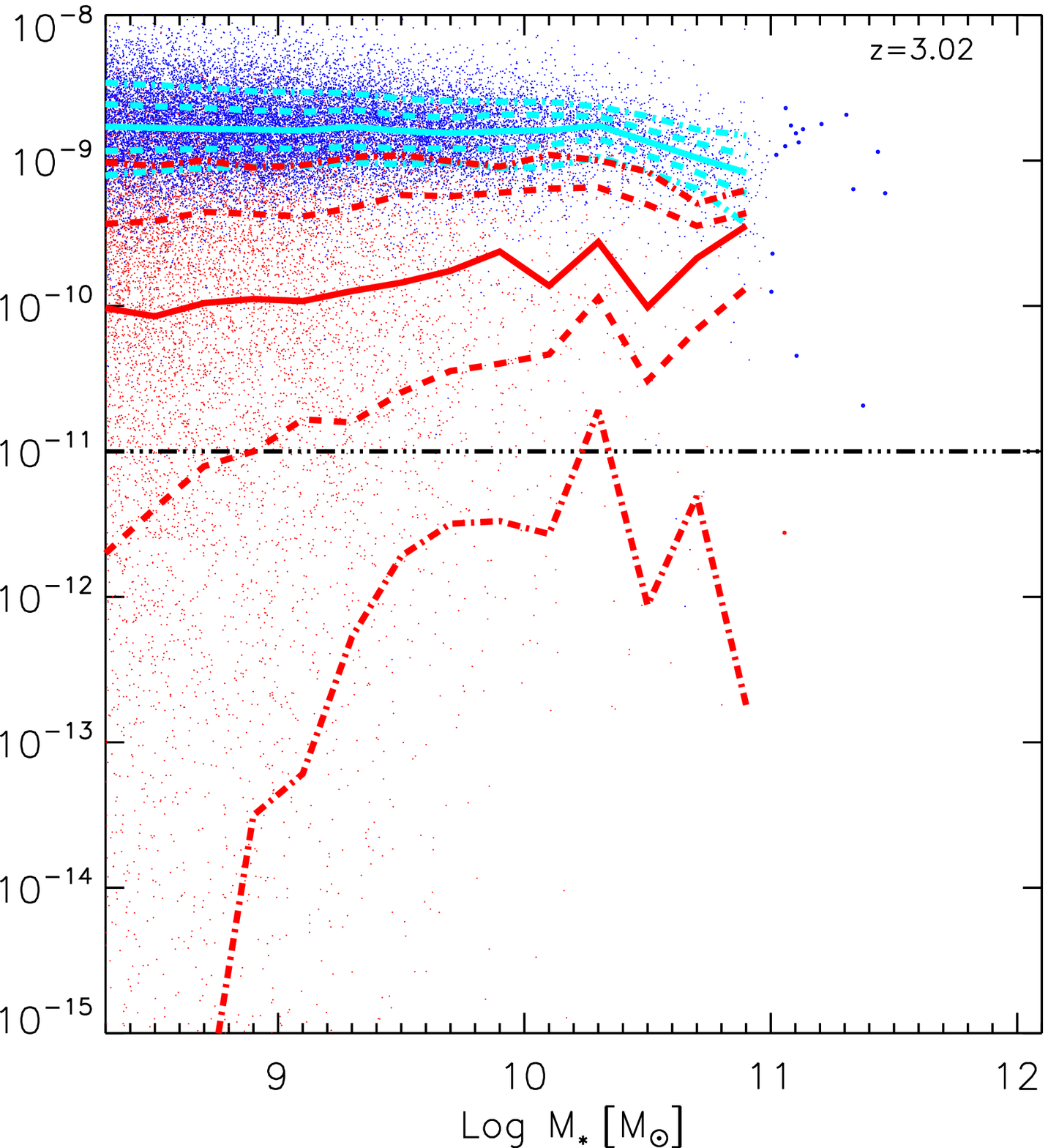} \\
\end{tabular}
\caption{Specific star formation rate-stellar mass relation at redshifts
  $z=0.79,1.61,2.16,3.02$ (different panels) for progenitors that are central
  (blue) and satellite (red) of passive galaxies in clusters at $z=0$. Solid
  lines represent medians, while dashed and dot-dashed lines represent the
  30th-70th and 15th-85th percentiles, respectively, cyan for centrals and red
  for satellites. Larger symbols have been used for galaxies more massive
    than $10^{11} \, M_{\odot}$.} The black horizontal dash-dotted line
  represents our threshold in specific star formation rate that separates
  passive and active galaxies.
\label{fig:ssfr_mass}
\end{figure*}
\end{center}

In this section we analyse in more details the galaxy population in
our proto-cluster regions, and how it evolves as a function of cosmic
time. In particular, we focus our attention on those galaxies that
will become passive at $z=0$. To this aim, we select all galaxies at
redshift $z=0$ that have $\log SSFR <-11$, which we consider as 
the threshold between passive and star-forming galaxies at any redshift. 
For this sample of galaxies, we select their progenitors in our proto-cluster 
regions and study when and how their star formation rate is suppressed.

We start by considering the location of all progenitors of passive galaxies
today in the specific star formation rate-stellar mass plane. This is shown in
Fig.~\ref{fig:ssfr_mass} for four different redshifts. Progenitors are
colour-coded according to their hierarchy (blue points and cyan lines are used
for central galaxies and red points and lines for satellites). At high
redshifts, 90 per cent of the progenitors are active, while the fraction of
active galaxies decreases with decreasing redshift with 9 per cent of the
central progenitors passive at $z\sim 0.8$, and 69 per cent of the satellites
being passive at the same redshift. Moreover, Fig.~\ref{fig:ssfr_mass}
  also shows that the most massive passive galaxies are centrals since 
  $z \sim 2.2$. Observational studies are typically limited to the most
  massive passive galaxies in proto-clusters, those with $M_* > 10^{10.5} \,
  M_{\odot}$. As we will discuss below, a large fraction of these galaxies in
  our model are quenched centrals, from AGN feedback. So, while the
  growth of the red-sequence at lower redshift is caused primarily by quenching
  of satellite galaxies, the passive galaxies that we can observe in distant
  clusters and in protoclusters are likely to be quenched by a different
  mechanism (AGN feedback).

\begin{figure}
\begin{center}
\includegraphics[scale=.57]{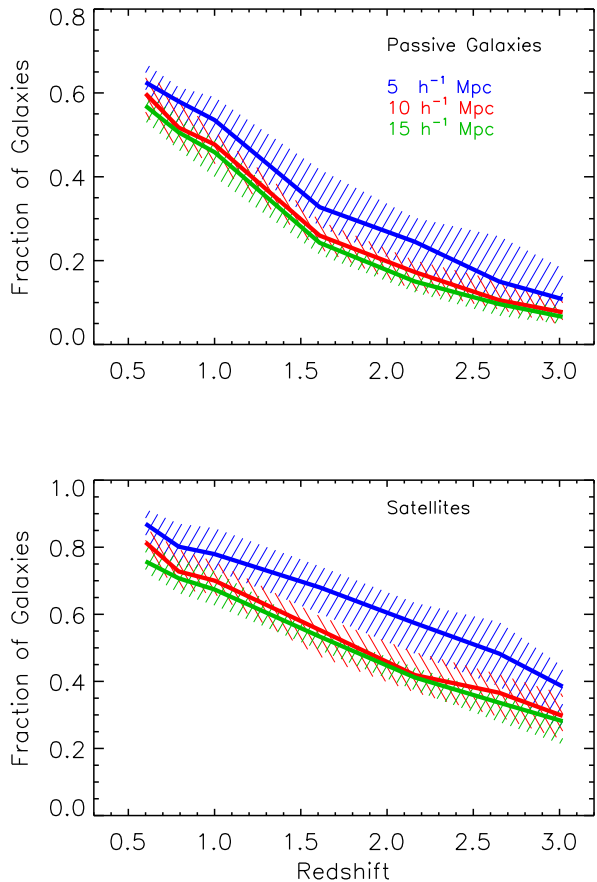} 
\caption{Fraction of passive galaxies (top panel) and satellites
  (bottom panel) within boxes of different sizes (different colours)
  centred around the CG, as a function of redshift. Solid lines represent the median value
  calculated on the 27 proto-cluster regions (22 for the largest box), 
  while dashed areas represent the distribution between the 20th and 80th
  percentiles. The threshold in stellar mass is $M_* = 10^9 \,
  M_{\odot}$, but the trends shown do not depend on the particular
  threshold chosen.}
\label{fig:redsat}
\end{center}
\end{figure}

\begin{figure}
\begin{center}
\includegraphics[scale=.46]{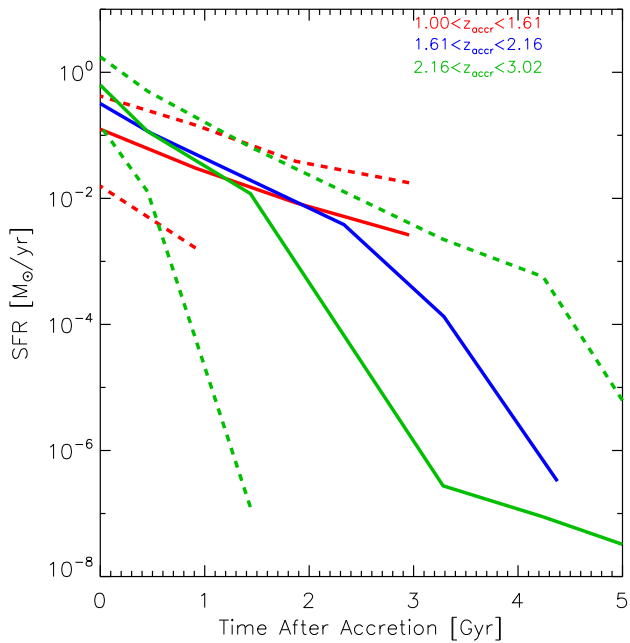} 
\caption{Star formation rate of progenitors of satellite galaxies with 
  mass in the range $[10^{9}-10^{9.5}] \, M_{\odot}$ accreted at different 
  redshifts, as shown in the legend, as a function of time elapsed since accretion. Solid lines 
  represent the median value calculated on the 27 proto-cluster regions 
  (22 for the largest box), while dashed areas represent the distribution 
  between the 20th and 80th percentiles (the scatter corresponding to the
    blue line is comparable to that obtained for progenitors accreted at higher
    redshift, and is omitted to make the figure less crowded).}
\label{fig:sfrhistory}
\end{center}
\end{figure}

In our model, galaxies that are accreted on larger structures
(i.e. become satellites) are instantaneously stripped of the hot gas
reservoir that can fuel new material available for star formation
through gas cooling. The combination of instantaneous hot gas
stripping and relatively efficient supernova feedback adopted makes
the transition from blue to red very short for satellite galaxies
(\citealt{simone06,wang07}). It is therefore expected that
most of the satellites in our models are
passive. Fig.~\ref{fig:redsat} shows the fraction of passive
progenitors in the top panel and that of progenitors that are
satellites in the bottom panel, as a function of the redshift of our
proto-cluster regions. We have considered in this case only
progenitors with stellar mass larger than $10^9 \, M_{\odot}$ and
residing in boxes with different size (different colours). Both
fractions depend on the size of the box, and decrease with
increasing size of the box because larger boxes capture more star
forming progenitors, preferentially located at larger distances from
the CG with respect to passive ones. At redshift $z \gtrsim 2$ the
fraction of passive progenitors is not higher than 30 per cent, which
confirms that the dominant population in proto-clusters is made up by
star forming galaxies.

Both the fraction of satellite progenitors and that of passive ones
increase at decreasing redshift, but the figure shows that the
fraction of passive progenitors is smaller (less than half) than the
fraction of satellites at high redshift, and increases faster towards
lower redshift. 

As highlighted above, the suppression of gas cooling makes satellite 
galaxies passive on relatively short time-scales, in the model. This 
is a well known problem pointed out by several authors 
(\citealt{weinmann06,font08,weinmann10,michaela12}).
As shown by Figure \ref{fig:redsat}, many satellites are still 
active at high-z. This is because they have been accreted very recently 
and the amount of cold gas available is still enough to keep them active. 

In Figure \ref{fig:sfrhistory} we show the star formation rate of progenitors
of satellite galaxies with mass in the range $[10^{9}-10^{9.5}] \, M_{\odot}$
accreted at different times (red, blue and green lines for $1<z_{accr}<1.61$,
$1.61<z_{accr}<2.16$ and $2.16<z_{accr}<3.02$, respectively), as a function of
the time after accretion.  This plot shows that progenitors accreted earlier
(green line) have the tendency to be slightly more star forming than
those accreted later (red line), at the time of accretion (although the scatter
is very large). We find that this is driven by a higher cold gas fraction for
galaxies accreted at higher redshift. In particular, we find that progenitors
accreted in the range $2.16<z_{accr}<3.02$ have gas fraction ($M_{cold}/M_*$)
larger by about 34 per cent than that of progenitors accreted in the range
$1.61<z_{accr}<2.16$, and about twice that of progenitors accreted in the
redshift range $1<z_{accr}<1.61$. Moreover, Figure \ref{fig:sfrhistory} 
  suggests that, at a given stellar mass, the quenching time-scale is shorter
  for galaxies accreted at high redshift (the green line is steeper than the
  red line), where galaxies tend to be less massive and to reside in lower halo
  mass and eject mass more efficiently. We have analysed galaxies having different 
  values of stellar mass measured at z=0, and verified that the qualitative picture 
  remains the same.

\begin{figure}
\begin{center}
\includegraphics[scale=.47]{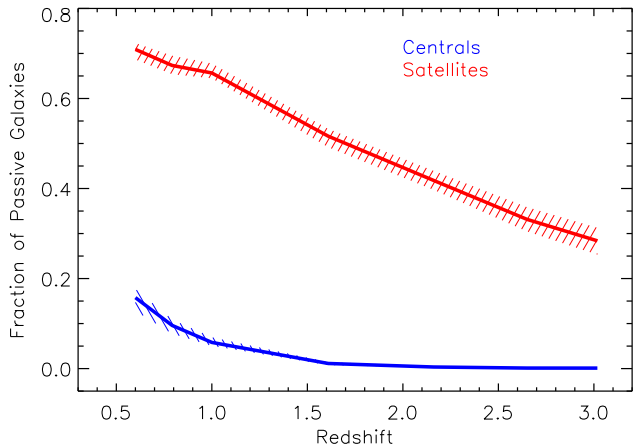} 
\caption{Fraction of passive satellites and centrals, progenitors of
  passive galaxies in clusters as a function of redshift. Solid lines
  represent the median value calculated on the 27 proto-cluster
  regions, while dashed areas represent the distribution between the
  20th and 80th percentiles.}
\label{fig:satcen_passive}
\end{center}
\end{figure}

Figure \ref{fig:redsat} is suggesting that satellite galaxies are 
likely the major contributors in building-up the passive-sequence.
In order to better quantify the relative contribution to the
passive-sequence given by satellites and centrals, we plot in Figure
\ref{fig:satcen_passive} the fraction of passive satellites (red) and
passive centrals (blue) that are progenitors of passive galaxies in
clusters and stellar mass larger than $10^9 \, M_{\odot}$, 
as a function of time. Focusing on the redshift range of
interest of proto-clusters ($z \gtrsim 2.$), we find that most of the
satellites are still active (as found above). Indeed, the fraction 
of passives is $\sim 0.4$ at $z=2$, and it increases with decreasing 
redshift, reaching 0.7 at $z \sim 0.6$. 
Different is the picture for centrals: they are almost all active down
to $z \sim 1$ and only 15 per cent of them are passive at $z \sim 0.6$.

A small, but not-negligible, fraction (around 15 per cent at $z \sim 0.6$) of
central galaxies that are progenitors of passive galaxies in clusters start to
be quenched at $z \sim 1.3$. We addressed this point and found that many are
intermediate-mass galaxies (30 per cent with $M_* \gtrsim 10^{10.4} \,
M_{\odot}$ and a median value of $M_* \gtrsim 10^{10.2} \, M_{\odot}$ at $z
\sim 0.6$).  The main responsible for their quenching is found to be a strong
AGN feedback, that prevents cooling of hot gas (at any stellar mass) which
would replenish the cold reservoir. Nevertheless, we find that around 10 per
cent of these galaxies experience a burst of star formation between $z=1$ and
$z=1.5$.  This is due to a rapid cooling of hot gas (coming from newly accreted
satellites), that enhances star formation. In a few cases, the burst is also
driven by mergers.

The analysis done in this section points out that galaxies in proto-clusters
are star forming objects at any time, and the passive-sequence of galaxies
emerges at around redshift $z \sim 1$. Centrals are star forming down to low
redshift, but a small fraction of them are quenched by AGN feedback, while
satellites are quenched by stripping of their hot gas reservoir at the
  time of accretion. The time during which they keep forming stars depends on
the amount of cold gas available at accretion.

\section{Conclusions}
\label{sec:conclusions}
We have analysed a sample of 27 proto-cluster regions extracted from a 
set of N-body simulations, that become massive clusters, with 
$M \sim 10^{15} \, M_{\odot}$, at $z=0$. These regions have been built 
by considering all progenitors of $z=0$ galaxies within the virial radius 
$R_{200}$ of the clusters into which these objects will evolve. 

The case study shows that progenitors of galaxies in massive galaxy clusters 
distribute in a very large region at high redshift. This region is dominated 
by central galaxies at high redshift. Their number decreases with 
time because many become satellites, clustering around the central object. We 
find that the velocity dispersion of galaxies increases with cosmic time, in 
good agreement with observations, and that of satellites increases faster, in 
line with a picture where galaxies grow in mass as centrals and then join the 
densest regions as satellites.

  In agreement with estimates based on the spatial distribution of galaxies
  around the Spiderweb Galaxy \citep{hatch09}, we find that $95 $ per cent of
  the satellites within a radius of $150 \, kpc$ from the central object at
  redshift $z \sim 2$ will merge with it by $z=0$. This implies that mergers
  are important in building-up the central object and for the overall evolution
  of such regions.

Our analysis highlights that proto-clusters are very extended objects. Indeed,
we find that at most 60 per cent of progenitors are located within a box of $15
\, \hm Mpc$ size centred on the central galaxy of the proto-cluster, at
redshift higher than $\sim 1.5$. The percentage decreases drastically in
smaller boxes, to at most 20 per cent for $5 \, \hm Mpc$ apertures.  This
demonstrates that one has to consider fairly large regions, having comoving
sizes larger than $15 \, \hm Mpc$, in order to trace the distribution of a
large fraction of the progenitors of galaxies belonging to local
clusters. Moreover, we find a shift between the geometrical centre of the
proto-cluster and the position of the central objects (that observers usually
take as the centre of the proto-cluster), that affects the fraction of
progenitors in the box. This suggests that attention must be payed when
comparing observations with model predictions when the size of the
proto-cluster region is relatively small. 

We find that the fraction of outliers, i.e. those galaxies that are not
progenitors of any cluster galaxy at $z=0$, is dependent on the galaxy stellar
mass and on redshift, and does not strongly depend on the size of the box up to
$15$ comoving Mpc. On average, we find that about 30 per cent of galaxies with
stellar mass smaller than $\sim 10^{10} \, M_{\odot}$, and about 20 per cent of
galaxies with larger stellar masses, are outliers in proto-cluster regions at
$z \sim 3$. Slightly smaller fractions are found at lower redshift and/or
considering only progenitors of galaxies within the virial radius (instead of
$2 \cdot R_{200}$). It is virtually impossible to distinguish outliers from
actual progenitors just by looking at their properties.


We have focused on progenitors of passive-sequence galaxies at $z=0$, and
studied when the passive-sequence emerges. At high redshift ($z \sim 3$) we
find that 90 per cent of progenitors are still active, confirming that star
forming galaxies are the dominant population in proto-clusters. Moreover,
central and satellite galaxies show a different evolution with time. We find
that satellites are the main contributors in building-up the passive sequence,
but many of them are still active at high redshift (see Fig.~11) and
their quenching is regulated by the amount of cold gas available at the time of
accretion.  Nevertheless, the timescale for quenching depends on the time of
accretion, being shorter for progenitors accreted at higher redshifts.  Central
galaxies contribute little and only at lower redshift. Only a small fraction of
them start to be quenched after $z\sim 1.3$, and a strong AGN feedback is the
responsible for their quenching, independently on the galaxy stellar mass.

Galaxies in proto-clusters are actively star forming, more intensively at
increasing redshift. Nevertheless, if we consider a minimum dust correction,
our model predicts proto-cluster regions that are systematically less star
forming than those observed, and tend to be even less star forming than
observed if we take into account apertures larger than $1 \, {\rm Mpc}$.
  Although large uncertainties exist in the data, the lowest observationally
  measured star formation rates quoted in this paper are still significantly
  higher than the highest predicted by our model.

\section*{Acknowledgements}
EC and GDL acknowledge financial support from the European Research Council
under the European Community's Seventh Framework Programme (FP7/2007-2013)/ERC
grant agreement n. 202781. This work has been supported by 973 program
(No. 2015CB857003, 2013CB834900), the NSF of Jiangsu Province (No. BK20140050),
the NSFC (No. 11333008 and No. 11550110182), the Strategic Priority Research
Program the emergence of cosmological structures” of the CAS (No. XDB09010403),
the PRIN-MIUR 201278X4FL Evolution of cosmic baryons funded by the Italian
Ministry of Research, by the PRIN-INAF 2012 grant “The Universe in a Box:
Multi-scale Simulations of Cosmic Structures”, by the INDARK INFN grant and by
“Consorzio per la Fisica di Trieste”.  Simulations have been carried out at the
CINECA National Supercomputing Centre, with CPU time allocated through an ISCRA
project and an agreement between CINECA and University of Trieste.  We
acknowledge partial support by the European Commissions FP7 Marie Curie Initial
Training Network CosmoComp (PITN-GA-2009-238356).  EC is also funded by Chinese
Academy of Sciences President's International Fellowship Initiative, Grant
No.2015PM054.  We thank the referee Alvaro Antonino Orsi for useful comments
that helped us improving our manuscript.

\label{lastpage}

\bibliographystyle{mn2e}
\bibliography{biblio}

\end{document}